\long\def\equalign
#1\end#2{\def\\{\cr\noalign{\smallskip}}
    \refstepcounter{equation}$$ \vcenter{\tabskip=5pt
    \halign{\hfil$\displaystyle{##}$&&$\displaystyle{{}##}$\hfil\cr
    #1 \cr}}\end{#2}}
 \def\theequation{\arabic{equation}}

\documentstyle[12pt]{article}

\newcommand \beq{\begin{eqnarray}}
\newcommand \eeq{\end{eqnarray}}
\begin{document}
\title{COLLECTIVE FERMIONIC EXCITATIONS IN SYSTEMS WITH A
LARGE CHEMICAL POTENTIAL}  \author{ Jean-Paul BLAIZOT  and
Jean-Yves OLLITRAULT\\
Service de Physique Th\'eorique, CE-Saclay\\
91191 Gif-sur-Yvette, France\\
}
\maketitle
\begin{abstract}
We study  fermionic excitations in a cold
ultrarelativistic plasma. We construct explicitly the
quantum states associated with the two branches which
develop in the spectrum as the chemical
potential  is raised. The collective nature of the long
wavelength excitations  is cleary exhibited.\end{abstract}
\newpage

\def\square{\hbox{{$\sqcup$}\llap{$\sqcap$}}}   
\def\grad{\nabla}                               
\def\del{\partial}                              

\def\frac#1#2{{#1 \over #2}}
\def\smallfrac#1#2{{\scriptstyle {#1 \over #2}}}
\def\half{\ifinner {\scriptstyle {1 \over 2}}
   \else {1 \over 2} \fi}

\def\bra#1{\langle#1\vert}              
\def\ket#1{\vert#1\rangle}              

\def\simge{\mathrel{%
   \rlap{\raise 0.511ex \hbox{$>$}}{\lower 0.511ex \hbox{$\sim$}}}}
\def\simle{\mathrel{
   \rlap{\raise 0.511ex \hbox{$<$}}{\lower 0.511ex \hbox{$\sim$}}}}


\def\parenbar#1{{\null\!                        
   \mathop#1\limits^{\hbox{\fiverm (--)}}       
   \!\null}}                                    
\def\nunubar{\parenbar{\nu}}
\def\ppbar{\parenbar{p}}


\def\buildchar#1#2#3{{\null\!                   
   \mathop#1\limits^{#2}_{#3}                   
   \!\null}}                                    
\def\overcirc#1{\buildchar{#1}{\circ}{}}


\def\slashchar#1{\setbox0=\hbox{$#1$}           
   \dimen0=\wd0                                 
   \setbox1=\hbox{/} \dimen1=\wd1               
   \ifdim\dimen0>\dimen1                        
      \rlap{\hbox to \dimen0{\hfil/\hfil}}      
      #1                                        
   \else                                        
      \rlap{\hbox to \dimen1{\hfil$#1$\hfil}}   
      /                                         
   \fi}                                         %


\def\subrightarrow#1{
  \setbox0=\hbox{
    $\displaystyle\mathop{}
    \limits_{#1}$}
  \dimen0=\wd0
  \advance \dimen0 by .5em
  \mathrel{
    \mathop{\hbox to \dimen0{\rightarrowfill}}
       \limits_{#1}}}                           

\def\real{\mathop{\rm Re}\nolimits}     
\def\imag{\mathop{\rm Im}\nolimits}     

\def\tr{\mathop{\rm tr}\nolimits}       
\def\Tr{\mathop{\rm Tr}\nolimits}       
\def\Det{\mathop{\rm Det}\nolimits}     

\def\mod{\mathop{\rm mod}\nolimits}     
\def\wrt{\mathop{\rm wrt}\nolimits}     


\def\TeV{{\rm TeV}}                     
\def\GeV{{\rm GeV}}                     
\def\MeV{{\rm MeV}}                     
\def\KeV{{\rm KeV}}                     
\def\eV{{\rm eV}}                       

\def\mb{{\rm mb}}                       
\def\mub{\hbox{$\mu$b}}                 
\def\nb{{\rm nb}}                       
\def\pb{{\rm pb}}                       

%
\def\journal#1#2#3#4{\ {#1}{\bf #2} ({#3})\  {#4}}

\def\AdvPhys{\journal{Adv.\ Phys.}}
\def\AnnPhys{\journal{Ann.\ Phys.}}
\def\EurophysLett{\journal{Europhys.\ Lett.}}
\def\JApplPhys{\journal{J.\ Appl.\ Phys.}}
\def\JMathPhys{\journal{J.\ Math.\ Phys.}}
\def\LettNuovoCimento{\journal{Lett.\ Nuovo Cimento}}
\def\Nature{\journal{Nature}}
\def\NPA{\journal{Nucl.\ Phys.\ {\bf A}}}
\def\NPB{\journal{Nucl.\ Phys.\ {\bf B}}}
\def\NuovoCimento{\journal{Nuovo Cimento}}
\def\Physica{\journal{Physica}}
\def\PLA{\journal{Phys.\ Lett.\ {\bf A}}}
\def\PLB{\journal{Phys.\ Lett.\ {\bf B}}}
\def\PhysRev{\journal{Phys.\ Rev.}}
\def\PRC{\journal{Phys.\ Rev.\ {\bf C}}}
\def\PRD{\journal{Phys.\ Rev.\ {\bf D}}}
\def\PRL{\journal{Phys.\ Rev.\ Lett.}}
\def\PhysRept{\journal{Phys.\ Repts.}}
\def\ProcNatlAcadSci{\journal{Proc.\ Natl.\ Acad.\ Sci.}}
\def\ProcRoySoc{\journal{Proc.\ Roy.\ Soc.\ London Ser.\ A}}
\def\RevModPhys{\journal{Rev.\ Mod.\ Phys.}}
\def\Science{\journal{Science}}
\def\SovPhysJETP{\journal{Sov.\ Phys.\ JETP}}
\def\SovPhysJETPLett{\journal{Sov.\ Phys.\ JETP Lett.}}
\def\SovJNuclPhys{\journal{Sov.\ J.\ Nucl.\ Phys.}}
\def\SovPhysDoklady{\journal{Sov.\ Phys.\ Doklady}}
\def\ZPhys{\journal{Z.\ Phys.}}
\def\ZPhysA{\journal{Z.\ Phys.\ A}}
\def\ZPhysB{\journal{Z.\ Phys.\ B}}
\def\ZPhysC{\journal{Z.\ Phys.\ C}}


\setcounter{equation}{0}

\section{Introduction}

The spectrum of fermion excitations in a hot relativistic plasma composed
of electrons, positrons and photons (or quarks, antiquarks and gluons)
presents
an interesting structure [1-8]. In particular, for small fermion
masses $m$ and momenta $p$, i.e. $m,p\simle gT$
where $g$ is the coupling constant and $T$ the temperature, it is split. One of
the two
branches becomes the ordinary fermionic branch at large momentum or large mass.
The
other has  features
 of a collective excitation, and for this reason has been sometimes dubbed the
``plasmino''
 \cite{plasmino}, by analogy with the plasmon oscillation. However, one should
keep in
mind that, when $p\simle gT$, both branches have a collective character and, as
we shall
see, the whole structure of the spectrum at small momentum reflects a
collective behavior
of the system. A recent calculation indicates, at least within the one-loop
approximation, that this structure in the spectrum does not occur as an abrupt
transition, but develops gradually as the temperature is raised \cite{BBS92}.

Similar effects are expected to occur at zero temperature, but large chemical
potential
$\mu$, i.e. when $g\mu\simge m,p$\cite{Pisarski89}. This
is the situation we explore in this paper. There are many physics motivations
for
studying the spectrum of fermions in very dense matter, ranging from the
physics of
dense stars, to that of heavy ion  collisions. These have stimulated recent
works\cite{Altherr92}. There are also current  investigations on more formal
questions
related  to QED with a chemical potential\cite{Polonyi91}. However, the  main
purpose
of the present paper is simply to get a better physical understanding of the
fermion
spectrum, especially at low momentum where collective phenomena develop. We do
this, in
particular, by providing an explicit calculation of the quantum states
corresponding to the fermionic excitations, which is possible at zero
temperature.

Some time ago,
Weldon presented a physical interpretation  for the splitting of the spectrum
at high
temperature,  based on an analysis of the quantum numbers of the
modes\cite{Weldon89a},
and an analogy with BCS theory of superconductivity\cite{Weldon89b} . However,
that did
not totally clarify the simple physics underlying the emergence of collective
fermionic
excitations in the high temperature plasma. In fact the analogy with BCS
superconductivity is misleading. The formation of Cooper
pairs involve particles localized in momentum space at the vicinity of the
Fermi
surface, and it produces an instability of the Fermi sea leading to a
reorganization
of the system as a whole. Here, as we shall see, the particles in the vicinity
of the
Fermi surface are only weakly perturbed, and the Fermi sea is stable.

We begin in the next section by summarizing some properties of the fermion
self-energy
at finite temperature and chemical potential, to leading order  in the coupling
strength $g$.  The spectrum of fermion excitations is presented, and some known
results
are recalled. We show in particular that, in the ultrarelativistic limit, most
properties of the spectrum depend on $\mu$ and $T$ only through the combination
$M^2\propto g^2(\mu^2+\pi^2T^2)$. In section 3, we discuss some general
properties of
the one-loop fermion self-energy at zero temperature and finite chemical
potential.
Analytic results are obtained in two simple cases, $m=0$ and $p=0$. We discuss
the
approach to the ultrarelativistic limit and the emergence of a
particle--antiparticle
symmetry at large $\mu$. We show that, as it is the case at finite
temperature, the structure in the spectrum of quasiparticles  develops
gradually as the
chemical potential is raised, due to the coupling of the quasiparticles to low
lying
states involving hard electrons and photons. We argue that the imaginary part
of the
self energy may not be interpreted as a damping of the quasiparticles.
 The main results of this paper are
contained in section 4 where we present a detailed calculation of the quantum
states
corresponding to the two branches in the excitation spectrum at small momentum.
The
collective nature of the two branches is clearly exhibited, and various
features of the spectrum are analyzed. Our conclusions are summarized in the
last
section which contains also a discussion of various aspects of the present
work, such
as its relevance to the case of finite temperature, and
also to QCD, the relation with kinetic
theory developed in \cite{QED}, and finally the relation with  a
 mode analogous to the plasmino, the ``plasmaron'' found in some calculations
of
the non relativistic electron gas\cite{Hedin67,Minnhagen74,Mahaux85}.

\def\bfgamma{\mbox{\boldmath$\gamma$}}
\def\bfalpha{\mbox{\boldmath$\alpha$}}
\def\bfsigma{\mbox{\boldmath$\sigma$}}
\section{Fermionic modes in ultrarelativistic plasmas}

In the presence of a chemical potential associated with the
conserved charge $Q=\int{\bar\psi\gamma_0\psi{\rm d}^3x}$, the Dirac
hamiltonian
$\bfalpha\cdot{\bf p}+m\gamma_0$ becomes
$\bfalpha\cdot{\bf p}+m\gamma_0-\mu$ (
$\bfalpha=\gamma_0\bfgamma$, $\gamma_0$ and $\bfgamma$ being the usual
 Dirac matrices). Thus, the net
effect of the chemical potential is to shift all the single particle energies
by $-\mu$.
Correspondingly, the  inverse of the free Dirac propagator may be written
\beq \label{g0}
G_0^{-1}({\bf p},
\omega)&=& -(\omega+\mu)\gamma_0+\bfgamma\cdot{\bf p}+m\nonumber\\
&=&(\epsilon_p-\mu-\omega)\gamma_0\Lambda^+_{\bf p}-
(\epsilon_p+\mu+\omega)\gamma_0\Lambda^-_{\bf p}
\eeq
where
\beq
\label{projector}
\Lambda^\pm_{\bf p}={1\over 2\epsilon_p}\left(\epsilon_p
\pm(\bfalpha\cdot{\bf p}+m\gamma_0)\right) \eeq
are the projectors on positive and negative energy solutions respectively
($(\Lambda^\pm)^2=
\Lambda^\pm$, $\Lambda^++\Lambda^-=1$),
$\epsilon_p=\sqrt{p^2+m^2}$, and $\omega$ is a complex variable whose
real part  measures a single particle energy with
respect to the chemical potential. In the following, we shall in fact absorb
$\mu$ into $\omega$, i.e. we substitute $\omega$ to $ \omega+\mu$ in
Eq.(\ref{g0})
so that
${\rm Re}\,\omega$ will measure the energy with respect to the
zero of energy in the absence of chemical potential. If needed, the
Feynman propagator is obtained as usual by letting $\omega$ approach the
real axis from above when ${\rm Re}\,\omega>\mu$, and from below when
${\rm Re}\,\omega<\mu$ \cite{ManyBody,BR86}.

For simplicity, we consider in the main text the interaction of a Dirac fermion
$\psi$ and
a massless scalar field $\phi$ with a Yukawa coupling $g\phi\bar\psi\psi$.
However, most of the discussion extends to QED (see Appendix B) and for this
reason  we shall often refer to the fermion as an ``electron'' and to the
scalar
particle as a ``photon''. In
  presence of the interaction, the inverse propagator is
  $G^{-1}=G^{-1}_0+\Sigma$, and, owing to
rotational invariance, the mass operator $\Sigma$ may be written in the form
\beq
\label{ABC}
\Sigma({\bf p},\omega )=a(p,\omega )\gamma^0+
b(p,\omega )\bfgamma\cdot{\bf p}+c(p,\omega )
\eeq
where $a,b$ and $c$ are three independent functions of the length
$p$ of the momentum, and the energy $\omega$.

To
order $g^2$ the mass operator at finite temperature and chemical
potential is the sum of two terms $\Sigma({\bf p},\omega)=\Sigma_1(
{\bf p},\omega)+\Sigma_0({\bf p},\omega)$, where  $\Sigma_0({\bf p},\omega)$ is
the
vacuum contribution, that is the mass operator for $T=\mu=0$, and $\Sigma_1(
{\bf p},\omega)$ is the correction due to finite $\mu$, finite $T$, effects.
The
latter is given by
 \beq \label{sigma1}\Sigma_1(
{\bf p},\omega)=g^2\int\frac{d^3k}
{(2\pi)^3}\frac{1}{2\omega_k} \biggl\{
\Lambda^+_{{\bf p}+{\bf k}}\frac{n_k-f^-_{{\bf p}+{\bf
k}}}{\omega-\epsilon_{{\bf p}+{\bf k}}-\omega_k}+
\Lambda^-_{{\bf p}+{\bf k}}\frac{f^+_{{\bf p}+{\bf
k}}+n_k}{\omega+\epsilon_{{\bf p}+{\bf k}}-\omega_k}
\nonumber\\
\qquad\qquad+\Lambda^+_{{\bf p}+{\bf k}}\frac{f^-_{{\bf p}+{\bf
k}}+n_k}{\omega-\epsilon_{{\bf p}+{\bf k}}+\omega_k}+
\Lambda^-_{{\bf p}+{\bf k}}\frac{n_k-f^+_{{\bf p}+{\bf
k}}}{\omega+\epsilon_{{\bf p}+{\bf k}}+\omega_k}
\biggr\}\gamma_0, \eeq
where
\beq
f^\pm_p=\frac{1}{e^{\beta(\epsilon_p\pm\mu)}+1}\qquad\qquad
n_k=\frac{1}{e^{\beta\omega_k}-1}
\eeq
are the occupation numbers respectively for the electrons ($f_p^-$),
the positrons ($f_p^+$), and the photons ($n_k$).
 The photon energy is $\omega_k=k$, except in
the calculation of the vacuum contribution (see below and Appendix A) where we
set
$\omega_k=\sqrt{k^2+\lambda^2}$, $\lambda$ playing the role of an infrared
cut-off. Note that in writing Eq.(\ref{sigma1}), we ignore
contributions involving possibly non--vanishing (in fact infrared divergent)
expectation values of $\bar\psi\psi$
(or $\bar\psi\gamma^0\psi$ in the case of QED). However,
such contributions to $\Sigma$ are independent of energy and momentum and
produce simply a constant shift in the single particle energies. Therefore,
they can
be absorbed in a redefinition of the chemical potential. Now, since this plays
no
role in the present discussion, we simply ignore these constant contributions.
(In the usual
treatment of the non-relativistic electron gas, one assumes that such
contributions are
cancelled by those arising from a positive background of positive charges. In
the case of QCD,
such contributions vanish because of color neutrality). Thus,  in zeroth order,
the chemical potential
will be taken equal to the Fermi energy, i.e.  $\mu=\sqrt{m^2+k_F^2}$, where
$k_F$ denotes the Fermi
momentum, related as usual to the density of the conserved charge.

The vacuum ($T=\mu=0$) contribution to the mass operator has a structure
similar to Eq.(\ref{sigma1}),
\beq
\label{sigma0}
\Sigma_0({\bf p},\omega)&=&g^2\int\frac{d^3k}
{(2\pi)^3}\frac{1}{2\omega_k}
\bigg\{
\Lambda^+_{{\bf p}+{\bf k}}\frac{1}{\omega-\epsilon_{{\bf p}+{\bf
k}}-\omega_k}+
\Lambda^-_{{\bf p}+{\bf k}}\frac{1}{\omega+\epsilon_{{\bf p}+{\bf k}}+\omega_k}
\bigg\}\gamma_0.\nonumber\\&&
\eeq
The integrations over momentum are diverging and require renormalization
(see Appendix A).
However, as we shall see shortly, in the ultrarelativistic
limit where $m\sim g\mu ,g\, T$, and for soft modes with $\omega, p\sim
g\mu,gT$,
the contribution of $\Sigma_0$ is of higher order in $g$
than  the dominant terms of $\Sigma_1$, and it will not be discussed
further in this section.

In the limit where $m,\,\omega,\, p\ll\mu,\, T$, the integrand
in Eq.~(\ref{sigma1})  contributes mostly in the region $
k\sim\mu\ {\rm or}\ T$ where  we can replace  $\epsilon_{{\bf p}+{\bf
k}}$ by $k+p\cos\theta$, with $\theta $ the angle between
${\bf p}$ and ${\bf k}$.  Furthermore,  the first and the last term in
Eq.~(\ref{sigma1}), for which the energy denominators are of order
$k\sim\mu, T$, are small compared to the second and third terms where the
energy denominators are approximately  $\omega\pm p\cos\theta\ll
T,\,\mu$. The same reasoning shows that the vacuum contribution, which has the
same denominators as the first and last terms of $\Sigma_1$, can be discarded
in
a first approximation. Keeping  only the dominant terms, one
obtains then
\beq \label{sigmaultra}
\Sigma({\bf p},\omega )&\approx&g^2\int\frac{d^3k}
{(2\pi)^3}\frac{1}{2\omega_k} \bigg\{
\Lambda^-_{{\bf p}+{\bf k}}\frac{f^+_{{\bf p}+{\bf
k}}+n_k}{\omega+\epsilon_{{\bf p}+{\bf k}}-\omega_k}+
\Lambda^+_{{\bf p}+{\bf k}}\frac{f^-_{{\bf p}+{\bf
k}}+n_k}{\omega-\epsilon_{{\bf p}+{\bf k}}+\omega_k}
\bigg\}\gamma_0 \nonumber\\&\approx&{g^2\over 16\pi^2} \int_0^\infty{k{\rm
d}k\left( f^-_k+ f^+_k+2n_k\right) } \int_{-1}^1{{d\cos\theta\over\omega
-p\cos\theta}\left( \gamma^0-{{\bfgamma\cdot {\bf p}} \over p}\cos\theta\right)
}
.\nonumber\\ & &
\eeq
 The integrals over
$k$ and over $\theta $ are decoupled and can be done exactly. Note that  the
 $k$-integration extends to the region of soft momenta for which the
approximation done is non longer
valid. Therefore, the expression (\ref{sigmaultra}) should be used only in
situations where the
weight of such soft momenta is negligible. A straightforward calculation gives
the following results
for the functions $a$, $b$ and $c$ defined in Eq.(\ref{ABC}):
\beq\label{abcultra} {\rm Re}\ a({\bf
p},\omega)&=&\frac{M^2}{2p}\ln\left| \frac{\omega+p}{\omega-p}\right|
\nonumber\\ {\rm Re}\ b({\bf
p},\omega)&=&\frac{M^2}{p^2}\left( 1-\frac{\omega}{2p}\ln\left|
\frac{\omega+p}{\omega-p}\right|\right) \nonumber\\
{\rm Re}\ c({\bf p},\omega)&=&0
\eeq
with $\omega$ real and
\beq\label{defM}
M={g\over 4\pi}\sqrt{\mu ^2+\pi ^2T^2}.
\eeq
In the ultrarelativistic limit, the self-energy $\Sigma$ depends on the
chemical potential and the temperature only through this parameter $M$.

The functions $a$ and $b$ have imaginary parts when $|\omega|<p$, given by
\beq\label{Landau}  {\rm Im}\ a(p,\omega)
&=& -\pi{M^2\over 2p}\nonumber\\ {\rm Im}\ b(p,\omega) &=& \phantom{-}\pi
{M^2\omega\over 2p^3 }
\eeq
${\rm if} -p<\omega<p$ and $0$ otherwise. The states  to which
the soft fermion couples, and which are responsible for this imaginary part,
may be associated to
virtual transitions where hard particles scatter on the soft one, with little
deflection, and a
possible change in their quantum numbers. For example, a hard electron can
annihilate on a soft
positron
 and ``turn  into'' a hard photon.  The kinematics of such processes
is somewhat analogous to
 that involved in the Bremsstrahlung of soft photons by fast electrons.
Here, the soft particles are electrons or positrons and the hard particles
are electrons, positrons or photons. These virtual transitions have
$\omega\sim p\cos\theta$ and their number is constant, independent of $p$; thus
their density increases as $1/p$
when $p$ decreases. In fact, this density is measured by the imaginary part of
$a\propto
M^2/2p$. Essentially all particles can contribute to these transitions, as
reflected in the
momentum integral  in Eq.~(\ref{sigmaultra}). Note
that when $p\to 0$, $b\to 0$, and $a(p,\omega)\to M^2/\omega$,
 so that, in this limit, ${\rm
Im}\,a(\omega)\propto \delta(\omega)$.
 The existence of an imaginary part in the region
 $\left\vert\omega\right\vert <p$ is reminiscent of the familiar  Landau
damping of electromagnetic
waves in ordinary plasmas.  Here, however, the imaginary part is of the same
order of magnitude
as the real part, i.e. ${\cal O}(M)$, and excitations with
such energies ($\left\vert\omega\right\vert <p$) would be  quickly
damped.

 Well defined quasiparticles exist for $|\omega|>p$. Their  energies are
the poles of the propagator, and are therefore given by the solutions of the
equation  $\det(G^{-1})=0$, where
\beq \label{gmoinsun}
G^{-1}({\bf p},\omega )=\left(
a-\omega\right) \gamma_0+ \left( 1+b\right) {\bfgamma}\cdot{\bf p} +(c+m),
\eeq
or
\beq\label{branches} \omega-a=\varepsilon\sqrt{ \left( 1+b\right) ^2{\bf
p}^2+(c+m)^2},
\eeq
with $\varepsilon=\pm 1.$
Using the explicit form of the functions $a$, $b$ and $c$
just derived, one can write Eq.(\ref{branches}) in the form
\beq\label{quasipart}
\frac{\omega^2-p^2-m^2}{p^2}= {\rm h}\left(\frac{\omega}{p}\right){\rm
h}\left(-\frac{\omega}{p}\right) \left({M\over p}\right)^4+2\left({M\over
p}\right)^2
\eeq
with
\beq {\rm h}(x)\equiv 1-{x+1\over 2}\ln\left\vert {x+1\over x-1}\right\vert .
\eeq The solution of Eq.~(\ref{quasipart}) gives $M/p$ as a function of
$\omega/p$,
from which  one obtains the dispersion relation $\omega(p)$ in parametric form
for a
given $M$.  Note that Eq.(\ref{quasipart}) is even in $\omega$, and its
solutions come in
pairs with opposite signs. Thus, in the ultrarelativistic limit, there
is an apparent particle-antiparticle symmetry
for the soft quasiparticles excitation energies.
This is at first sight surprising
since such a symmetry is explicitly broken by the presence of the chemical
potential. We discuss this point further in the next section. We note here that
if $\omega$ is a
solution of Eq.~(\ref{quasipart}) with $\varepsilon=1$, then $-\omega$ is a
solution with
$\varepsilon=-1$. Furthermore, for a given $\varepsilon$, there are two
solutions of opposite signs.
We call  $\omega_+$ and $\omega_-$ the two  solutions corresponding to
$\varepsilon=1$,
with $\omega_+>0$, $\omega_-<0$.
We call ``normal'' the branches $\varepsilon\omega_+$  and ``abnormal'' the
branches
$\varepsilon\omega_-$. Thus, for
$\omega >0$, the branch $\omega_+$ is the normal one, the branch $-\omega_-$
the
abnormal one. This seemingly complicated notation will be fully justified in
section 4 after we have clarified the origin of the observed pattern of states.
As we shall see,
 $\varepsilon$ can be identified with the eigenvalue of $\gamma_0$ in the case
where
$p=0$, and with that of $\bfalpha\cdot\hat{\bf p}$ when $m=0$ ($\hat{\bf
p}={\bf p}/p$).

The spectrum is displayed in Figs.1. If $m$ is small enough compared to $M$,
the group velocity on the abnormal positive branch, $-{\rm
d}\omega_-/{\rm d}p$, is negative  for small $p$, and $-\omega_-(p)$  presents
a
minimum for a finite value of the momentum.
The condition under which  this
phenomenon occurs,  $m/M<\sqrt{14/(3\sqrt{5}) -2}\approx 0.295$,
can be derived easily by expanding  Eq.(\ref{quasipart}) in
powers of $p/\omega$.
For $m=0$, we have
\beq\label{groupvel}
\omega_\pm (p) =\pm M\left[1\pm{1\over 3}{p\over M}+\dots\right].
\eeq
Thus, in particular, the slopes of the two branches $\omega_\pm$ are the same
at small $p$,
$d\omega_\pm/dp=1/3$ when $m=0$.  As $M$ becomes smaller than  $m$, the
normal branch $\varepsilon\omega_+(p)$
goes over to the normal fermion dispersion relation
$\varepsilon\sqrt{p^2+m^2}$, while the abnormal
branch $\varepsilon\omega_-(p)$ approaches
$-\varepsilon p$. The asymptotic behavior
for $p\gg M$ and $m=0$, is  \beq
\omega_+(p)&\sim &p+{M^2\over p}\nonumber\\
\omega_-(p)&\sim & -p\left[ 1+\exp(-2p^2/M^2)\right]
\eeq

The residues at the quasiparticle poles can also be calculated. They have
simple
expressions in the two limiting cases where $p=0$ or $m=0$. Then, the
projectors
(\ref{projector}) reduce to $(1\pm\gamma_0)/2$ when $p=0$, and to
$(1\pm\bfalpha\cdot\hat{\bf p})/2$ when $m=0$,  and we can write
, in the appropriate subspace,
\beq G({\bf
p},\omega)\gamma_0=\frac{1}{(\varepsilon\epsilon_{\bf
p}-\omega)+\Sigma_\varepsilon(\omega)}
\approx\frac{z_\pm}{\varepsilon\omega_\pm-\omega} \eeq
where $\varepsilon$ denotes the eigenvalue, $\pm 1$, of $\gamma_0$ in the case
${\bf
p}=0$, or that of $\bfalpha\cdot\hat{\bf p}$ when $m=0$,
$\Sigma_\varepsilon\equiv a+\varepsilon
c$ when $p=0$, and $\Sigma_\varepsilon\equiv a+\varepsilon bp$ when $m=0$. The
residues are given by
$z_\pm^{-1}=1-\del\Sigma_\varepsilon/\del\omega|_{\omega_\pm}$
and they depend only on
the character of the mode, normal or abnormal, irrespective of the sign of
$\omega$, i.e. of $\varepsilon$. When $m=0$, \beq\label{residm0}
z_\pm={\omega_\pm^2-p^2\over 2M^2} . \eeq When $p=0$, \beq\label{residp0}
z_\pm={1\over
1+M^2/\omega_\pm^2}. \eeq
  The variations with $p$ of the residues are  displayed in Figs.1.
When $m\not= 0$, the residue of the
abnormal branch is always smaller than that of the  normal branch. It becomes
negligible when either $m\gg M$ or $p\gg M$. For $p=0$ we have $z_++z_-=1$.
For $p\not= 0,$
$z_++z_-<1$; this is because the spectral weight has a smooth component
between
$\omega=-p$ and $\omega=p$.  We will comment on
this in section 4.

\section{One loop results at $T=0$}
\def\kF{k_F}

We now study the full one loop correction to the fermion propagator
at $T=0$ and $\mu\not= 0$.
In this case, the mass operator has a simple analytic expression when either
$p=0$ or $m=0$. The corresponding formulae are
collected in Appendix B.
Some of the discussion in this section parallels that
in Ref.\cite{BBS92}, where a similar study was presented at finite temperature
and
zero chemical potential.
It is instructive to see how the structure in the spectrum develops as the
chemical
potential is raised and the ultrarelativistic limit is approached.
However, one should keep in mind that the full one--loop
calculation is not entirely consistent, as revealed for example by the
order of magnitude of some contributions to the imaginary part. We shall
discuss the
physical meaning of these imaginary parts and see why it is not always
meaningful to
associate those in the vicinity of the quasiparticle poles with the damping of
quasiparticles.

\subsection{The mass operator at $T=0$}

At zero
temperature, there are no bosons, i.e. $n_k=0$, nor
antifermions, i.e. $ f^+_p=0$, and all single particle states with energies
less than
$\mu$ are occupied, i.e. $ f^-_p=\theta(\mu-\epsilon_p)$. The mass operator
takes the form
\beq \label{fullsigma}
\Sigma({\bf p},\omega)=g^2\int\frac{d^3k}
{(2\pi)^3}\frac{1}{2\omega_k}\bigg\{ &\Lambda^+_{{\bf p}+{\bf k}}&
\frac{1-f^-_{{\bf p}+{\bf k}}}{\omega-\epsilon_{{\bf p}+{\bf k}}-
\omega_k}\nonumber\\
+&\Lambda ^+_{{\bf p}+{\bf k}}&
\frac{f^-_{{\bf p}+{\bf k}}}{\omega-\epsilon_{{\bf p}+{\bf k}}+
\omega_k}+\Lambda^-_{{\bf p}+{\bf k}}\frac{1}{\omega+\epsilon_{{\bf p}+{\bf
k}}+\omega_k}
\bigg\}
\gamma_0,\nonumber\\
\eeq
where we have kept together the vacuum contribution and the finite chemical
potential correction, without writing explicitly the
renormalization counterterms (see Appendix A for details).
Note that the imaginary part, which we discuss below, is finite and not
affected by the
renormalization. In order to facilitate the foregoing discussion, we label by
(i), (ii)
and (iii) the three contributions to $\Sigma({\bf p},\omega)$ in
Eq.~(\ref{fullsigma}).
These are illustrated by the diagrams of Figure 2. In these diagrams, the
labels on the
lines correspond to the  momentum and the energy of the single particle state
to which the line
refers. A fermionic line going upwards represents an electron above the Fermi
surface. A
fermionic line going downwards represents a hole in the Fermi sea if the energy
is
positive (case (ii)), and a positron (a hole in the Dirac sea) if the energy is
negative
(case (iii)). Thus, reading the diagram (ii) in Fig.2 from bottom to top, one
finds a hole
excitation with energy $-\omega$ and momentum $-{\bf p}$ which makes a virtual
transition
to a state containing a photon with momentum ${\bf k}$ and energy $k$, and a
hole in the
single particle state with momentum  ${\bf p+k}$ and energy $\epsilon_{\bf
p+k}$.

It is perhaps useful at this point to recall that the poles of the Green's
function
occur for values of $\omega$ equal to $\pm (E_{\bf p}^{N\pm 1}-E_0^N)$, where
$E_{\bf p}^{N\pm 1}$ is the energy of a state with $N\pm 1$ particles, and
total
momentum ${\bf p}$, while $E_0^N$ is the energy of the ground state with $N$
particles.
(Note that in this discussion we count particles with respect to the fully
occupied Dirac
sea, and similarly for the energies.) Perturbative calculations are meaningful
as long as
$E_0^N-E_0^{N-1}<\mu<E_0^{N+1}-E_0^N$ \cite{BR86}. Values of $\omega$ such that
${\rm Re}\,
\omega>\mu$ correspond to possible energies of states with $N+1$ particles,
while
values of $\omega$ such that ${\rm Re}\,
\omega<\mu$ correspond to the negative of  energies of states with $N-1$
particles. For
example, the state obtained by removing a particle of momentum ${\bf p}$
in the unperturbed Fermi sea has a momentum $-{\bf p}$, and an energy $-\omega=
E_{\bf p}^{N-1}-E_0^N=-\epsilon_{\bf p}$, with $\omega<\mu$. The same reasoning
applies to
positrons considered as holes in the Dirac sea. Removing an electron in the
Dirac sea
in a single particle state of momentum ${\bf p}$ and energy $-\epsilon_{\bf
p}$, produces a
state with $N-1$ particles, a momentum $-{\bf p}$, and an energy
$-(-\epsilon_{\bf
p})=\epsilon_{\bf p}$.

The mass operator acquires imaginary parts in several energy domains where the
energy denominators
in Eq.~(\ref{fullsigma}) vanish for $\omega$ real. To discuss these, we note
the
following inequalities \beq
\label{useful}
\epsilon_{{\bf p+k}}+k\ge\epsilon_{\bf p},\qquad\qquad
0\le\epsilon_{{\bf p+k}}-k\le\epsilon_{\bf p}.
\eeq
It is then easily seen that, quite generally, the imaginary part of (i)
vanishes when
$\omega<\mu$. This follows from the fact that the occupation factor in the
numerator
forces $\epsilon_{{\bf p+k}}>\mu$, so that the denominator
$\omega-\epsilon_{\bf p+k}-k$
cannot vanish if $\omega<\mu$. Similarly, the occupation factor $f^-_{\bf p+k}$
in the
numerator of (ii) forces $\epsilon_{{\bf p+k}}<\mu$ so that the imaginary part
of (ii)
vanishes when $\omega>\mu$. Finally, using the first of the inequalities
(\ref{useful})
one gets that the imaginary part of  (iii) vanishes when $\omega>-m$.

 Now, the precise
domains in which the imaginary part does not vanish depend on $p$. These are
illustrated
in Figure 3. The last term of Eq.~(\ref{fullsigma}), (iii), contributes an
imaginary part
when $\omega<-\epsilon_{\bf p}$, as easily deduced by using the first of
inequalities
(\ref{useful}). This corresponds physically to the process by which a positron
(off its
energy shell) with energy $|\omega|$  and momentum  ${\bf p}$   decays
into a positron of momentum ${\bf p} +{\bf k}$  and a photon of momentum ${\bf
-k}$, on
their respective mass shells; this process may occur in the vacuum, for example
under the
action of a weak perturbation which displaces the positron slightly off its
mass shell,
and it is not affected by  the Fermi sea of electrons. We shall refer to the
continuum
of states into which a positron can decay as the positron-photon continuum. It
corresponds to region (III) in Fig.3.

The first term in Eq.~(\ref{fullsigma}), labelled (i),
has, for $p>k_F$, an imaginary part when $\omega>\epsilon_{\bf p}$. This
imaginary part
lies in the vicinity of the electron mass shell, and reflects the possible
decay of an
electron above the Fermi surface  into another electron with less energy and a
photon. As
in the positron case just discussed, this process may also  occur in the
vacuum; however,
here, the presence of the Fermi sea eliminates some of the states available for
the  decay
in free space, because of the Pauli principle reflected by the occupation
factor
$1-f_{{\bf p}+{\bf k}}$ in Eq.~(\ref{fullsigma}). In the case $p<k_F$, this
factor
actually reduces the energy domain allowed by the kinematics, i.e.
$\omega>\epsilon_{\bf p}$, to
$\omega>\mu+k_F-p$,  and indeed leads to a cancellation of the imaginary part
in the
region $\epsilon_{\bf p}<\omega<\mu+k_F-p$. Then, for $p<k_F$,  the imaginary
part of (i) is non zero in a
domain ($\omega>\mu+k_F-p$) which can be far from $\epsilon_{\bf p}$ and it is
not immediately
related to a possible  decay of the single particle excitation. The continuum
of states
 discussed  in this paragraph will be referred to as the electron-photon
continuum. It
corresponds to region (I) in Fig.3.

The existence of states of the $N+1$ particle  system with momentum less than
the Fermi
momentum $k_F$ reflects modifications of the Fermi sea due to the interactions.
Because
of these, the single particle states are neither fully occupied below the Fermi
surface,
nor completely empty above it. These  modifications of the Fermi sea are
implicitly taken into account, to order $g^2$, in the expression
(\ref{fullsigma}) of the
mass operator. The interplay between the single particle properties and these
ground state
corrections  is perhaps best understood by considering the energy shift of a
single
particle state obtained in second order perturbation theory. This is obtained
 by solving Dyson's equation for the pole of the propagator,  setting
$\omega=\epsilon_{\bf p}$ in the mass operator $\Sigma({\bf p},\omega)$. The
first term in
$\Sigma({\bf p},\epsilon_{\bf p})$, i.e. (i), gives an attractive contribution,
while the
second term, i.e. (ii),  gives a repulsive contribution.
 This is easily seen by using
inequalities (\ref{useful}): the denominator of (i) is always negative, while
that
of (ii) is always positive.  To understand this better, consider   a particle
state above
the Fermi surface, i.e. with $p>k_F$, and  recall
that the poles of the Green's function correspond,  for $\omega>\mu$, to the
energy difference $E^{N+1}_{\bf p}-E^N_0$, where $E^{N+1}_{\bf p}$ is the
energy of the
system with $N+1$ particles and momentum $\bf p$. In carrying out the second
order
calculation of this energy difference, one is lead to write the total
correction as
 $\Delta \epsilon_{\bf p}+\Delta E^{N+1}_0-\Delta E^N_0$. The first term,
$\Delta
\epsilon_{\bf p}$, is the correction obtained from (i); it is negative and can
indeed be
interpreted as the second order correction to the energy of the single particle
state
${\bf p}$, due to its coupling to states with one electron (above the Fermi
surface) and
one photon. On the other hand, the quantity $\Delta E^{N+1}_0$ represents a
correction to
the the energy of the Fermi sea, taking into account the presence of the extra
particle.
The full second order correction is given by $\Delta E^N_0$, and is negative.
Because of
the Pauli principle, the presence of the extra particle  suppresses  some of
the
intermediate states (those having a particle with momentum ${\bf p}$ above the
Fermi surface) involved in the full second order correction to $E_0^N$; this is
illustrated in Fig. 4. As a result, $\Delta E^{N+1}_0>\Delta E^N_0$, and the
contribution
of (ii), which can be identified to $\Delta E^{N+1}_0-\Delta E^N_0$, is
positive. A similar interpretation may be given for states below the Fermi
surface,
exchanging the role of contributions (i) and (ii).

We come now to the second term in Eq.~(\ref{fullsigma}), labelled
(ii). In contrast to (i) and (iii) where vacuum contributions explicitly enter,
this term
has no counterpart in the vacuum and is entirely due to the Fermi sea of
electrons.  It
contains the  dominant contribution in the ultrarelativistic limit. It has for
$p<k_F$ an
imaginary part in the region $\mu-k_F-p<\omega<\epsilon_{\bf p}$. This
imaginary part measures
the density of states with momentum $\bf p$ containing  one hole and one photon
with
respect to the unperturbed Fermi sea. Such states may couple with a hole at the
bottom
of the Fermi sea, or  a hole at the top of the Dirac sea, i.e. with a positron.
A pictorial
representation of all these states, which will play a central role in our
discussion (in particular in section 4), is given in Fig 5. One may view the
corresponding virtual transitions as processes by which a hole in the Fermi
sea, or a positron, gets ``filled up'' by electrons falling into it from above
in the Fermi sea, emitting at the same time a photon.  If $p>k_F$, (ii) has an
imaginary part in the domain $\mu-p-k_F<\omega<\mu-p+k_F$. The existence of
states of the $N-1$ particle system having a momentum greater than the Fermi
momentum reflects, similarly to the case  discussed above, a modification of
the Fermi sea due to interactions. The continuum of states corresponding to
the virtual transitions described in this paragraph will be refered to as the
hole-photon continuum. This correspond to region (II) of Fig.3.

In the ultrarelativistic limit, the hole-photon continuum is entirely contained
in the
region $|\omega|>p$ (see section 2, the discussion after Eq.(\ref{Landau}).
However,  the above discussion shows that the imaginary part does not
strictly  vanish outside this interval. The additional contribution, in the
interval  $p<\omega <\epsilon_{\bf p}$, can be calculated  assuming $\mu\gg
\omega,p,m$ and $\omega-p\gg g^2\mu$. The result is,   \beq\label{imultra}
{\rm Im}\ a &=& -\pi {g^2\over 32\pi^2} {\omega\left[
m^4-(\omega^2-p^2)^2\right]
\over (\omega^2-p^2)^2}\nonumber\\
{\rm Im}\ b &=& -\pi {g^2\over 32\pi^2} { m^4-(\omega^2-p^2)^2
\over (\omega^2-p^2)^2}\nonumber\\
{\rm Im}\ c &=& -\pi {g^2\over 32\pi^2} {2m\left[ m^2-(\omega^2-p^2)\right]
\over\omega^2-p^2}
\eeq
The same expressions hold also, up to a global  change of sign of ${\rm Im}\,
b$,  in the
region $\omega <-\epsilon_{\bf p}$ correponding to the positron-photon
continuum.
When $p$, $m$ and $\omega$ are of order $g\mu$,
${\rm Im}\,\Sigma({\bf p},\omega)$, as given by Eq.(\ref{imultra}),
is  of order $g^3\mu$,
i.e. smaller by two powers of $g$ than when
$|\omega|<p$. This comes from the fact that almost all the states in the Fermi
sea
contribute to the hole-photon continuum, while only a restricted phase space is
involved in
the calculation leading to Eqs.(\ref{imultra}). Indeed  the phase space
vanishes as
$\omega$ approaches $\epsilon_{\bf p}$, as may be seen on these expressions
(\ref{imultra}). Note that Eqs.(\ref{imultra}) do not involve the chemical
potential.
This is to be contrasted to the case
$T\not= 0$, where
the phase space  depends explicitly on the temperature through the statistical
factors.
Thus at  $T\not= 0$ the imaginary part of the one-loop mass operator is of
order $g^2 T$ (instead of $g^3\mu$ here), because the number of soft photons
of energy $\sim gT$ is of order $1/g$.

We have seen in section 2 (see also Fig.1) that, in the ultrarelativistic
limit, there are
four quasiparticles of a given momentum $p<k_F$, whose energies are the
solutions of
Eq.(\ref{branches}). Some of these quasiparticles appear in regions where ${\rm
Im}
\Sigma$ is small but non vanishing.  One may then be tempted to associate
these small
imaginary parts  to a weak damping of the  quasiparticles. Because of them, the
quasiparticle energies are indeed shifted to complex values,
$\omega=\omega_\pm-i\gamma_\pm$, where $\gamma_\pm$ is always positive
($\gamma_\pm=-z_\pm {\rm Im}\Sigma(\omega_\pm)$).
(It is interesting in this respect to notice that without the
vacuum contribution, there would be some imaginary part for
$\epsilon_{\bf p}<\omega<\mu$ coming from the term labelled (i), with
the wrong sign.)
 The normal quasiparticle branch with
$\varepsilon=1$ always  satisfies $\omega_+ >\epsilon_{\bf p}$ and thus has no
imaginary part
(the branch $\omega_+$ is above the hole-photon continuum). On the other  hand,
the normal
branch $-\omega_+<-\epsilon_{\bf p}$ is always damped (it is always in the
positron-photon
continuum). As to the abnormal branches, there are two  possibilities. Either
$-\omega_->\epsilon_{\bf p}$ (case $M\gg m$ in Fig.1); then the branch
$-\omega_-$ is above
the  hole-photon continuum, and the branch $\omega_-<0$ is inside the
positron-photon
continuum. Or $-\omega_-<\epsilon_{\bf p}$ (case $M\approx m$ in Fig.1), and
then it is the mode
$-\omega_->0$ which is damped, being found inside the hole-photon continuum.
Now one may
question whether the fact that a branch of the spectrum lies inside a continuum
of states
implies necessarily a damping of the corresponding single particle excitation.
When $M$ is sufficiently large compared to $m$, possible damping would occur
because of a
mixing of the single particle excitation with the positron-photon continuum .
However, a  consistent  calculation of the damping should take into account the
fact that the edge of this continuum, taken here to be the unperturbed positron
energy $\epsilon_{\bf p}$, is strongly modified by the interactions, which
affect actually   both the propagation of the fermions and that of the photons.
Taking as a minimal approximation the modification of the fermion energy in
the calculation of the damping, one would obtain that the damping is forbiden
by the kinematics (the branch $-\omega_+$ becomes in this approximation the
edge of the continuum). The difficulty  encountered here is identical to that
noted in a related context (the damping of gluonic excitations in a
quark-gluon plasma) by several authors\cite{Kal80,Heinz87,Pisarski88}.
Technically, it finds its origin in  the fact that the imaginary parts, as
obtained from the one loop approximation, being of order $g^3 \mu$, are
inconsistent since higher loop diagrams can contribute to the same order.
Consistent calculations require resummations as explained in
Refs.\cite{Braaten90a}.

In order to get further insight into the properties of the quasiparticles,
and the characteristic features of the ultrarelativistic limit, we now briefly
consider successively the particular cases where $p=0$ and $m=0$.

\subsection{Massive fermion at rest}

The equation which  determines the poles of the propagator (in the region where
the
imaginary part is small), namely Eq.(\ref{branches}),  becomes, when $p=0$
\beq \label{mmodes} \omega-\varepsilon m =
{\rm Re}(a_1+\varepsilon c_1) ,
\eeq
where we have ignored the vacuum contribution, and $\varepsilon$ denotes the
eigenvalue of $\gamma^0$.
Analytical expressions for ${\rm Re}\, a_1$ and
${\rm Re}\, c_1$, as well as for the corresponding imaginary parts can be found
in
Appendix B. The quantity ${\rm Re}(a_1+\varepsilon c_1)$ is plotted in Figs.6,
as a
function of $\omega/m$ for
various values of $\mu/m$. In general, when
$\mu\not= 0$, there is no  symmetry under the transformation $\omega\to
-\omega$,
$\varepsilon\to -\varepsilon$.
However, this symmetry emerges at large $\mu$, as is already
apparent in Figs.6 (the function ${\rm Re}\, a_1$ becomes
odd in $\omega$ and ${\rm Re}\, c_1$ becomes negligible at large $\mu$).
A graphical solution of Eq.~(\ref{mmodes}) is obtained by taking the
intersection  of the curve ${\rm Re}(a_1+\varepsilon c_1)$ with the straight
line $\omega-\varepsilon m$. One sees that the normal unperturbed pole at
$\omega =\varepsilon m$  is shifted to the right $(\varepsilon=1)$ or to the
left $(\varepsilon=-1).$ Furthermore, when $\varepsilon=1,$  two additional
poles appear on both sides of
 the logarithmic divergence at
$\omega=\mu-k_F,$
even if $\mu$ is
arbitrarily small. When $\varepsilon=-1,$ a similar situation occurs with
two new poles
slightly above $\omega=\mu-k_F,$ but only if $\mu$
is large enough. The physical relevance of these extra solutions may be
asserted by
considering the imaginary part of $a+\varepsilon c$ (see Eq.~(\ref{imfini})).
One sees that the imaginary part remains large in the vicinity of the pole
closest and above $\mu-k_F.$ This pole correspond to an excitation which is
killed by Landau damping. The other two poles correspond to physical
excitations with negligible widths. As we have seen earlier, it is consistent
to ignore the small imaginary parts in the ultrarelativistic limit. A
consistent calculation of the damping rates of the quasiparticles requires
going beyond the one loop apparoximation.

The structure of the self--energy can be understood from the
dispersion relation
\beq
\label{dispersion}{\rm Re}\,\Sigma(\omega)={\rm P}\int{{d\omega'\over
2\pi}{{\rm
Im}\,\Sigma(\omega')\over \omega'-\omega}}
\eeq
where the symbol P in front of the integral indicates that the principal
value is to be taken.  The region contributing to the dominant imaginary
part is the  small $\omega$ region where the imaginary part goes as
$\omega^{-3}$. This behavior reflects the increase of the phase space for
processes in which a hole at the bottom of the Fermi sea couples with holes
near the Fermi surface and photons with momentum $\sim k_F$. When $\mu$ grows,
the lower bound of the allowed phase space $\mu -k_F\sim m^2/(2\mu)\to 0$, and
the number of such transitions  becomes increasingly important. Thus, at large
$\mu$, the imaginary part is approximately given by ${\rm Im}\Sigma\sim
-(g^2m^4/32\pi\omega^3)$. Using the dispersion relation  (\ref{dispersion}),
and keeping only the dominant contribution coming, when $\omega\sim g\mu$,
from the lower bound of the integral $\sim m^2/(2\mu)$, one gets ${\rm
Re}\,\Sigma\sim M^2/\omega={\rm Re}\, a(0,\omega )$ ($M=g\mu/4\pi$), in
agreement with
Eq.(\ref{abcultra}). Thus the leading behavior of ${\rm
Re}\,\Sigma$ is determined by the hole-photon continuum, to which all electrons
in the Fermi sea can contribute. This explains the emergence of the symmetry
$\omega\to -\omega$ in the ultrarelativistic limit. Most of the particles
involved in the virtual transitions making the hole-photon continuum are hard
particles. In pictorial terms, an electron with momentum $p\sim \mu\gg m$
falling into a hole with momentum $p\sim g\mu$, does not ``see'' the mass gap,
i.e. it does not feel the difference between a hole at the bottom of the Fermi
sea
and a hole at the top of the Dirac sea.

In the high density  limit Eq.~(\ref{mfini}) has
only two solutions  for either value of  $\varepsilon$  (the third solution,
slightly above $\mu-k_F,$ disappears in this limit), that is to say four
solutions. The positive solutions are \beq
\label{solmfini}
\pm\omega_\pm=\pm{m\over
2}+\sqrt{M^2+\left(  {m\over 2}\right)^2}.
\eeq
Eq.(\ref{condit}) shows that the imaginary part vanishes for both
positive solutions if $M>m\sqrt{2}$, and only for the solution
$\varepsilon=1$ if $M<m\sqrt{2}$. As to the negative solutions, $-\omega$ has
no imaginary part when $\omega$ has one and vice-versa. As already
mentioned, this imaginary part is small and can be ignored.

\subsection{Massless fermion with a finite momentum}

When $m=0$, Eq.(\ref{branches}) becomes (as in the previous case we ignore the
imaginary
parts and the vacuum contribution)  \beq\label{pmodes}
\omega-\varepsilon p={\rm Re}\,(a_1+\varepsilon b_1p)
\eeq
and  $\varepsilon$ denotes the eigenvalue
of $\bfalpha\cdot\hat{\bf p}$ ($\hat{\bf p}={\bf p}/p$).
The exact expressions for the real and the imaginary parts of  $a_1$ and $ b_1
p$ are given in appendix B,  and ${\rm Re}(a_1+\varepsilon b_1 p)$ is
displayed in Figs.7  as a function of $\omega.$ One sees that the imaginary
part of $a+\varepsilon bp$ is continuous everywhere except at $\omega
=-\varepsilon p,$
as can be checked in Eq.~(\ref{impfini}). This
discontinuity at the edge of the hole-photon continuum is
an effect of the zero mass.
The imaginary part is significant only if $|\omega |<p$.
The number of processes
contributing to it is an increasing function of $\mu$.  We also note
that, as in the previous section, the real part of $a_1+\varepsilon b_1p$
is, in general, not
 odd under the changes of sign of both $\varepsilon$ and $\omega$. However,
it becomes so in the high $\mu$ limit, for reasons which have already been
explained.  Eq.~(\ref{pmodes}) can be solved graphically in Fig.7, exactly
in the same way as Eq.~(\ref{mmodes}) in Fig.6.
One sees that for a small value of
$g,$ Eq.~(\ref{pmodes}) always has two solutions in addition to the solution
close to $\omega=\varepsilon p.$ These new poles are on both sides of
$\omega=-\varepsilon p$, and the pole which is inside the hole-photon
continuum is Landau damped if $\mu\gg p$. Thus we have finally one new physical
pole for either value of $\varepsilon $ when  $m=0$, as we had for $p=0$.

Finally, let us mention that the states in the vicinity of the Fermi surface
are not much affected by the interactions in this order. This is
illustrated in Figs.8.  The scale on these figures shows indeed that $\Sigma$
is a small quantity when $p\sim k_F$ (it is completely perturbative, i.e. of
order $g^2$). The lines $\omega\mp p$ are almost vertical lines in  these
plots, indicating that single particle states are shifted from  their
unperturbed positions by negligible amounts. As for the imaginary  part, it
vanishes for $\omega=\mu$ as is evident on Eq.(\ref{impfini}).

We end this section by summarizing the physical picture which emerges from
this analysis. As we have seen, most of the interesting physics is due to the
hole-photon continuum. When $\mu$ is large, this continuum involves virtual
transitions in which hard electrons ``fall'' into holes at the bottom of
the Fermi sea, or at the top of the Dirac sea (soft positrons). The number of
such transitions grows as $\mu^2$, reflecting the fact that most particles of
the Fermi sea participate and suggesting the collective character of the
quasiparticles. We shall now complete this picture by presenting an explicit
calculation of the quantum states corresponding to the quasiparticles.

\def\bfSigma{\mbox{\boldmath$\Sigma$}}

\section{Explicit calculation of the states}

In this section we present an explicit construction of the quantum states
associated with each branch of the excitation spectrum. We shall work mainly
in the ultrarelativistic limit. As we shall see, many of the features which
were left unexplained in the previous analysis will find here a natural
interpretation.

We decompose the fermion and scalar fields into plane waves normalized in a
volume $\Omega$ \beq \label{Dirac} \psi({\bf
x})&=&\frac{1}{\sqrt{\Omega}}\sum_{\lambda,{\bf p}} e^{i{\bf p}\cdot {\bf
x}}\, w_{\lambda,{\bf p}}b_{\lambda,{\bf p}}\\ \label{Phi} \phi({\bf
x})&=&\frac{1}{\sqrt{\Omega}}\sum_{\bf k} \frac{1}{\sqrt{2\omega_k}}\left(
e^{i{\bf k}\cdot {\bf x}}\,a_{\bf k}+ e^{-i{\bf k}\cdot {\bf x}}\,
a^\dagger_{\bf k} \right) \eeq In the decomposition of the fermion field
$\psi({\bf x})$, we do not separate at this stage positive and negative energy
solutions of the Dirac equation, and  consider a positron  as a hole in the
sea of  negative energy electrons.
 In Eq.(\ref{Dirac}), we therefore sum over  positive  and negative energy
solutions and
 the   index $\lambda$  takes four values, two for the spin and two for the
sign of energy. This convention allows us to treat the Dirac and Fermi seas on
the same footing: all single particle states with energy
 less than $\mu$ are occupied in the unperturbed ground state $\ket{\Phi_0}$.
The fermion spinors are normalized according to $w^\dagger_{\lambda,{\bf p}}
w^{ }_{\lambda',{\bf p}'}=\delta_{\lambda,\lambda'}\delta_{{\bf p},{\bf p}'}$
and they obey the free Dirac equation  \beq  \label{freeDirac}
\left[\bfalpha\cdot{\bf p}+\gamma_0 m\right]w_{\lambda,{\bf p}}
=\epsilon_{\lambda,{\bf p}}w_{\lambda,{\bf p}}  \eeq The interaction
hamiltonian is  \beq H_{int}&=&g\int{\phi({\bf x})\bar\psi({\bf x})\psi({\bf
x})}d^3{\bf x}\nonumber\\ &=&\frac{g}{\sqrt{\Omega}}\sum_{{\bf p},{\bf
k},\lambda,\nu}\frac{1}{\sqrt{2\omega_k}} \bar w_{\nu,{\bf p}+{\bf
k}}w_{\lambda,{\bf p}}
 b^\dagger_{\nu,{\bf p}+{\bf k}}b_{\lambda,{\bf p}}\left( a_{\bf k}
+a^\dagger_{-{\bf k}} \right) \nonumber\\ \eeq

We shall show that, in  the ultrarelativistic limit, the quantum states
corresponding to soft
fermionic excitations can be obtained
 by diagonalizing the total hamiltonian $H=H_0+H_{int}$, where $H_0$ is the
free hamiltonian for electrons and photons,
 in the following basis of states
\beq\label{states}
 b_{\lambda,{\bf p}}\ket{\Phi_0}\qquad\qquad b_{\nu,{\bf p}+{\bf
k}}a^\dagger_{\bf
k}\ket{\Phi_0} ,\eeq
 where $\epsilon_{\lambda,{\bf p}}<\mu$, $\epsilon_{\nu,{\bf p}+{\bf k}}<\mu$
and, as said earlier, $\ket{\Phi_0}$ is the unperturbed ground state with
energy $E_0$. All the states (\ref{states}) have momentum ${\bf -p}$. The
first state is a hole with momentum $-{\bf p}$.  The second  represents a
state composed of a hole with momentum $-{\bf p}-{\bf k}$  accompanied by a
photon with momentum ${\bf k}$. These are eigenstates of  $H_0$ with
eigenvalues $E_0-\epsilon_{\lambda,{\bf p}}$ and   $E_0-\epsilon_{\nu,{\bf
p}+{\bf k}}+\omega_k$ respectively. A pictorial representation of these
excitations is given in Fig.5. The single particle energy
$\epsilon_{\lambda,{\bf p}}$ is, as mentioned above,  positive  or negative
depending on $\lambda$, namely  $\epsilon_{\lambda,{\bf p}}=\epsilon_p$ for
$\lambda=1,2$ and
 $\epsilon_{\lambda,{\bf p}}=-\epsilon_p$ for $\lambda=3,4$.

We call $X_{\lambda,{\bf p}}$ and  $Y_{\nu,{\bf p},{\bf k}}$,   respectively,
the amplitudes associated with the two types of states (\ref{states}) in the
decomposition of the eigenstates $\ket{\Psi_{-\bf p}}$ of $H$ on the
unperturbed
basis
\beq
\ket{\Psi_{-\bf p}}=\sum_\lambda X_{\lambda,{\bf p}}
b_{\lambda,{\bf p}}\ket{\Phi_0}+\sum_{{\bf k},\nu}Y_{\nu,{\bf p},{\bf k}}
b_{\nu,{\bf p}+{\bf k}}a^\dagger_{\bf
k}\ket{\Phi_0}.
\eeq
Note that in restricting ourselves to the states (\ref{states}) we are
performing a truncation of the Hilbert space. In particular, the states with
one hole and one photon are coupled to states involving two photons, two holes
and one electron above the Fermi sea. This truncation is in line with the high
density limit and is legitimate since the neglected states would contribute in
higher order in $g$ to the properties of the single hole state that we
are interested in. The eigenvalue problem
$(H_0+H_{int})\ket{\Psi_{-\bf p}}=-\omega\ket{\Psi_{-\bf p}}$
then takes the form  \beq
\label{eigenst}  &&(\omega-\epsilon_{\lambda, p})X_{\lambda,{\bf p}} -g
\sum_{{\bf k},\nu}\frac{1}{\sqrt{2\omega_k\Omega}}
\bar w_{\nu, {\bf p}+{\bf
k}}w_{\lambda,{\bf p}}  Y_{\nu,{\bf p},{\bf k}} =0\nonumber\\
&&-\frac{g}{\sqrt{2\omega_k\Omega}}\sum_\lambda \bar w_{\lambda,{\bf
p}}w_{\nu, {\bf p}+{\bf k}} X_{\lambda,{\bf p}}
+(\omega-\epsilon_{\nu,{\bf p}+{\bf k}}+\omega_k)Y_{\nu,{\bf p},{\bf k}}=0
\eeq where $-\omega$ measures the energy of a state with one particle less
than $|\Phi_0\rangle$ with respect to $E_0$ (see the discussion at the
beginning of section 3).  In the first equation, the sum runs over all
single particle states such that $\epsilon_{\nu,{\bf p}+{\bf k}}<\mu$.

 In order to calculate the eigenvalues $\omega$, we  use the second
equation to eliminate the amplitudes $Y_{\nu,{\bf p},{\bf k}}$.  One gets,
after a straightforward calculation,   \beq \label{elim1}
(\omega-\epsilon_{\lambda,{\bf p}})X_{\lambda,{\bf p}}-
\sum_{\lambda'}X_{\lambda',{\bf p}}\bar w_{\lambda',{\bf p}}M({\bf
p},\omega) w_{\lambda,{\bf p}}=0, \eeq where $M({\bf p},\omega)$ is
given by \beq\label{newsigma} M({\bf p},\omega)=g^2\int\frac{d^3k}
{(2\pi)^3}\frac{1}{2\omega_k}\bigg\{ \frac{\Lambda^+_{{\bf p}+{\bf k}}f^-_{{\bf
p}+{\bf k}}}{\omega-\epsilon_{{\bf p}+{\bf k}}+\omega_k}
+\frac{\Lambda^-_{{\bf p}+{\bf k}}}{\omega+\epsilon_{{\bf p}+{\bf
k}}+\omega_k} \bigg\}
\gamma_0.
\eeq
Here $\epsilon_{{\bf p}+{\bf k}}=|\epsilon_{\nu,{\bf p}+{\bf k}}|$ and
$\Lambda^\pm_{\bf p}$, introduced in Eq.(\ref{projector}) projects
onto states with positive or negative energy, i.e.
 $\Lambda_{\bf p}^+= \sum_{\nu=1,2} w_{\nu,{\bf p}} w_{\nu,{\bf p}}^\dagger$,
and $\Lambda_{\bf p}^-= \sum_{\nu=3,4} w_{\nu,{\bf p}} w_{\nu,{\bf p}}^\dagger$
(see
Eq.~(\ref{projector})). The occupation factor $f^-_{{\bf p}+{\bf k}}$ is 1 for
a particle in the
Fermi sea and 0 otherwise. The expression (\ref{newsigma}) coincides  with the
sum of the terms
(ii) and (iii) in the expression  (\ref{fullsigma}) of the mass operator
 $\Sigma(
{\bf p}, \omega)$. The term (i) in (\ref{fullsigma}) accounts, for small
$p$ and $\omega$, for modifications of the Fermi sea, as discussed in the
beginning of section 3. Such modifications are negligible in the
ultrarelativistic limit, and indeed they are not taken into account by the
eigenvalue problem (\ref{eigenst}). We shall come back to these at the end
of this section.
 The last term of (\ref{newsigma}), which involves positron intermediate
states, is also negligible in the ultrarelativistic limit, and we shall
consequently ignore it in the following.  By multiplying Eq.~(\ref{elim1})
by $\bar w_{\lambda,{\bf p}}$ and summing over $\lambda$, one easily gets
\beq
\label{eqG-1}
\gamma_0G^{-1}({\bf p},\omega)
\sum_\lambda X^*_{\lambda,{\bf p}}w_{\lambda,{\bf p}}=0.
\eeq
The values of $\omega$ for which Eqs.~(\ref{eigenst}) are satisfied are
therefore
the zeros of $G^{-1}$, that is, as announced earlier,  they coincide with the
poles of the Green function calculated here in the ultrarelativistic limit.

Using results established in section 2, we can write
Eq.~(\ref{eqG-1}) as follows
\beq
\label{intDirac}
(\omega-a)\chi=\left[(1+b)\bfalpha\cdot{\bf p} + m\gamma_0\right]\chi,
\eeq
where $\chi=\sum_\lambda X^*_{\lambda,{\bf p}}w_{\lambda,{\bf p}}$. The
eigenvalues are given by Eq.~(\ref{branches}). Let us recall that there
are two pairs of solutions which depend on the parameter $\varepsilon=\pm 1$,
with the normal solutions  such that $\omega\varepsilon>0$.  Besides, each
solution possesses a
two-fold spin degeneracy since the spin operator  ${\bfSigma}=\gamma_5\bfalpha$
commutes with both  $\bfalpha\cdot{\bf p}$ and $\gamma_0$.

The spinors $w_{\lambda,{\bf p}}$ are eigenstates of the free Dirac hamiltonian
(see
Eq.~(\ref{freeDirac})). In either case, $p=0$ or $m=0$, one can choose the
solution of Eq.~(\ref{intDirac}) to be one of these, i.e. choose
$\chi=w_{\lambda,{\bf
p}}$. Namely, when $m=0$, we can choose $\chi$ to be an eigenstate of
$\bfalpha\cdot\hat{\bf p}$ with eigenvalue $\varepsilon$. When $p=0$, we can
choose
$\chi$ to be an eigenstate of $\gamma_0$ with eigenvalue $\varepsilon$. In the
general case
where neither $m$ nor $p$ vanish,  the presence of the factor $(1+b)$ in
Eq.~(\ref{intDirac})
induces a mixing which depends on $\omega$ and which needs to be calculated
explicitly for each
eigenvalue. This will be done later, and we shall consider first these
particular cases where
$m=0$ or $p=0$. Then, for each eigenvalue, only one amplitude, say
$X_{\lambda,{\bf p}}\, $,
is non vanishing. Setting $X_{\lambda,{\bf p}}=1$,  i.e. deferring the
normalisation of the
states to the end of the calculation, we obtain the following simple expression
for the
amplitudes $Y_{\nu,{\bf p},{\bf k}}$ \beq \label{Yampl} Y_{\nu,{\bf p},{\bf
k}}=\frac{g}{\sqrt{2k\Omega}}\frac{\bar w_{\lambda,{\bf p}} w_{\nu,{\bf k}}}
{\omega-p\cos\theta}\, ,\eeq where we have used the fact that $k\gg p,m$ to
simplify the
denominator, and $\theta$ is the angle between ${\bf k}$ and ${\bf p}$.

In order to calculate explicitly the overlap
$\bar w_{\lambda,{\bf p}} w_{\nu,{\bf k}}$, we choose the chiral representation
of the
$\gamma$ matrices in which the spinors take the form
 \beq
\label{chiralpw}
u_{\lambda,{\bf p}}&=&\frac{1}{2\sqrt{\epsilon_p(m+\epsilon_p)}}
\left( \matrix{ (\epsilon_p+m+\bfsigma\cdot{\bf p})\phi_{\lambda\hat{\bf p}}\cr
(-\epsilon_p-m+\bfsigma\cdot{\bf p})\phi_{\lambda\hat{\bf p}}}
\right)\nonumber\\
v_{\lambda,{\bf p}}&=&\frac{1}{2\sqrt{\epsilon_p(m+\epsilon_p)}}
\left( \matrix{ (\epsilon_p+m-\bfsigma\cdot{\bf p})\phi_{\lambda\hat{\bf p}}\cr
(\epsilon_p+m+\bfsigma\cdot{\bf p})\phi_{\lambda\hat{\bf p}}} \right)
\eeq
where  $\hat{\bf p}={\bf p}/p$.
These spinors are eigenstates of helicity with helicity $\lambda$ if the two
components spinors are, i.e. if $({\bfsigma}\cdot\hat{\bf p})\phi_{\lambda\hat
p}=\lambda\phi_{\lambda\hat{\bf p}}$. In the limit $m=0$ they become
eigenstates of
$\gamma_5$, with eigenvalues $\lambda$ for $u_\lambda$ and $-\lambda$ for
$v_\lambda$. In the limit $p=0$ they are eigenstates of $\gamma_0$ with
eigenvalues $+1$ for $u$ and $-1$ for $v$. Note that in both cases the  spinor
$u_{\lambda,{\bf p}}$ is associated to $\varepsilon=1$ while the spinor
$v_{\lambda,{\bf p}}$
is associated to $\varepsilon=-1$.
 Since we now distinguish explicitly
positive and negative energy solutions, the  index $\lambda$ takes only two
values, i.e. $u_{1,-1}=w_{1,2}$, $v_{1,-1}=w_{3,4}$. Accordingly, we change the
decomposition of the fermion field into
\beq
\psi({\bf x})&=&\frac{1}{\sqrt{\Omega}}\sum_{\lambda,{\bf p}}
\left( e^{-ip\cdot x}\, u_{\lambda,{\bf p}}b_{\lambda,{\bf p}}
+e^{ip\cdot x}\, v_{\lambda,{\bf p}}d^\dagger_{\lambda,{\bf -p}}\right) ,
\eeq
where $d^\dagger_{\lambda,{\bf -p}}$ denotes the creation operator for a
positron with helicity $\lambda$ and momentum ${\bf -p}$. Also, from now
 on, $\ket{\Phi_0}$ denotes the Fermi sea (i.e. excluding the Dirac sea).

Let us first consider the case $m=0$. Then we have \beq
\bar u_{\lambda,{\bf p}} u_{\nu,{\bf k}}=\phi^\dagger_{\lambda\hat{\bf
p}}\phi_{\nu\hat k}
\delta_{\nu,-\lambda}
\qquad\qquad
\bar v_{\lambda,{\bf p}} u_{\nu,{\bf k}}=-\lambda\phi^\dagger_{\lambda\hat{\bf
p}}\phi_{\nu\hat
k}
\delta_{\nu,\lambda}.
\eeq
Only spinors with opposite chirality contribute in the coupling
to a scalar field.
 From these relations, one immediately deduces the form of the
eigenstates
\beq
\label{statem0}
\ket{\Psi_{-\bf p}^u}_{\pm,\lambda}&=&\sqrt{z_\pm}\left( b_{\lambda,{\bf p}}+
g\sum_{\bf k}\frac{1}{\sqrt{2\omega_k\Omega}}
\frac{\phi^\dagger_{\lambda\hat{\bf p}}\phi_{-\lambda\hat k}}{\omega_\pm-p\
{\rm
cos}\theta} b_{-\lambda,{\bf p}+{\bf k}}a^\dagger_{\bf k}\right)\ket{\Phi_0}
\nonumber\\
\ket{\Psi_{-\bf p}^v}_{\pm,\lambda}&=&\sqrt{z_\pm}\left(
d^\dagger_{\lambda,-{\bf p}}+ g\sum_{\bf k}\frac{1}{\sqrt{2\omega_k\Omega}}
\frac{-\lambda\,\phi^\dagger_{\lambda\hat{\bf p}}\phi_{\lambda\hat k}}
{-\omega_\pm-p\ {\rm cos}\theta}
b_{\lambda,{\bf p}+{\bf k}}a^\dagger_{\bf k}\right)\ket{\Phi_0} ,
\eeq
where $\ket{\Psi_{-\bf p}^u}$ (resp. $\ket{\Psi_{-\bf p}^v}$) is the eigenstate
corresponding  to $\varepsilon=+1$ (resp. $\varepsilon=-1$), the subscript $+$
(resp. $-$)
refers to the normal (resp. abnormal) branch, and $z_\pm$ is a normalization
constant. The
norm of the states can be calculated directly from  Eq.(\ref{statem0}) by
replacing the
sum over ${\bf k}$ by an integral,  as in Eq.(\ref{newsigma}), and noting
that  $\left|\phi^\dagger_{\lambda\hat{\bf p}}\phi_{\lambda\hat k}\right|={\rm
cos}(\theta/2)$ and $\left|\phi^\dagger_{\lambda\hat{\bf p}}\phi_{-\lambda\hat
k}\right|={\rm sin}(\theta/2)$. One then obtains
$z_\pm=(\omega^2_\pm-p^2)/2M^2$, which coincides with  the residues given by
Eq.(\ref{residm0}).  It may also be verified by an explicit calculation, using
the results of
section 2, that all the states are orthogonal.

One sees on Eqs.~(\ref{statem0}) that, as a result of the interactions, the
``bare'' hole $b_{\lambda,{\bf p}}\ket{\Phi_0}$ and the ``bare'' positron
$d^\dagger_{\lambda,-{\bf p}}\ket{\Phi_0}$ couple {\em independently} to a
  a coherent  superposition of hole-photon states. There is no
direct mixing between the hole and the positron states. Such a mixing does in
fact occur when $m\ne 0 $ and $p\ne 0$, as we shall see later, but even then,
this is not a dominant feature. The basic mechanism at work is best
understood by concentrating on what happens to one of the single particle
states, the hole state for example, as the interaction is switched on. This
is illustrated in Fig.7 where, on the left hand side is drawn the dispersion
relation $\omega(p)$ for the hole state in absence of interaction, while in
the right hand side is the split dispersion relation $\omega_\pm(p)$ which
results when the interaction is taken into account. The shaded area
represents the phase space occupied by the continuum of hole-photon states to
which the hole
couples. We have represented only the dominant part of the density of
states. As we have seen in section 3.1, the density of state is non
vanishing in the region $p<\omega<\epsilon_{\bf p}$, but it does vanish
as $\omega\to \epsilon_{\bf p}$. As it is clear from this figure, this
coupling to the continuum shifts some of the single particle strength from
the positive energy, where it normally belongs, to the negative energies
where it appears as an ``abnormal'' branch. In the limit where $m,p\ll
\mu$, the strength becomes equally distributed among the two branches. The
branches corresponding to the positrons are symmetrical to those just
described for the hole. The interpretation of the two branches with
positive $\omega$ is then clear. The normal branch is that part of the
electron strength which is pushed up by the interaction with the continuum;
the  abnormal branch is that part of the positron strength which is  pushed
up to positive energies by the same interactions. We note that the energy
shift, being of order $g\mu$ rather than $g^2\mu$, is non perturbative.
Most of the electrons in the Fermi sea contribute to this
phenomenon, and indeed the structure of the states just described has many
features of collective excitations in many body systems.

We  now discuss some properties of the spectrum in the light of these
remarks. When $m=0$ and  $p\to 0$,
$\omega\rightarrow M$ (see Eq.~(\ref{groupvel})) and $z_\pm\rightarrow 1/2$
(see Eq.~(\ref{residm0})) and all four states have a
similar structure, i.e. the single particle strength is equally distributed
among the four of them. When $p$ increases, the abnormal branch quickly
becomes $\omega_-\simeq -p$ with a residue $z_-\to 0$, while the normal branch
goes over to the usual fermion dispersion relation with $z_+\to 1$. At the same
time, the relative weight of the hole-photon component, $(1-z_\pm)/z_\pm$
decreases for the
normal branch, but increases for the abnormal one. As revealed by a simple
analysis, when
$p\simge M$, only the states
  at the edges of
the hole-photon continuum contribute to the abnormal branches. These states are
of the form $b_{\lambda,{\bf p+k}}a^\dagger_{\bf k}\ket{\Phi_0}$, with ${\bf
p}$
and ${\bf k}$ antiparallel when $\omega\sim -p$,  and  with
${\bf p}$ and ${\bf k}$ parallel when $\omega\sim p$.

 The two states (\ref{statem0}), i.e. the dressed hole and the dressed
positron,
have the same fermionic charge $-1$ with respect to the unperturbed ground
state.
This may be verified explicitly by taking the expectation values of the
operator
$Q\equiv\int :\psi^\dagger\psi: d^3{\bf x}=
\sum_{\lambda,p}\left(b_{\lambda,p}^\dagger b_{\lambda,p}
-d^\dagger_{\lambda,p}
d_{\lambda,p}\right)$ in either state. One may write this expectation value as
\beq Q_\pm=-(z_\pm+(1-z_\pm))\eeq where the first contribution is the
``direct''
contribution, i.e. that of the bare hole (or the bare positron), while
the second contribution is the ``induced'' part, that is the charge carried by
the
hole-photon component. When $p$ grows, $z_+\to 1$ and $z_-\to 0$, as already
mentioned. Thus, as $p$ grows,  on the normal branch the direct component
grows,
while on the abnormal branch it is the induced component which grows. Thus,
when $p\simge M$,
the single particle state completely decouples from the abnormal
branch.

One gets another illustration of this phenomenon by calculating the expectation
value of the  current  operator ${\bf I}=\int\psi^\dagger\bfalpha\psi d^3{\bf
x}=
\sum_{\lambda,p}\hat{\bf p}\left(b_{\lambda,p}^\dagger b_{\lambda,p}
-d^\dagger_{\lambda,p} d_{\lambda,p}\right)$. In a bare hole state
$b_{\lambda,{\bf p}}\ket{\Phi_0}$  this is ${\bf I}=-\hat{\bf p}$,  opposite
to that of the bare positron with the same momentum, i.e.
$d^\dagger_{\lambda,-{\bf p}}$. This difference may be understood by noting
that,
viewed as a hole in the Dirac sea, the positron is associated to a single
particle state whose energy, $-\epsilon_{\bf p}$  decreases with increasing
momentum ${\bf p}$, whereas the single particle state  corresponding to
the hole in the Fermi sea has an energy  $+\epsilon_{\bf p}$ which increases
with increasing momentum. This difference translates into a difference in the
signs of the velocities $d\epsilon_p/dp$ of the two excitations.  Taking the
expectation value of the current operator in the states(\ref{statem0}) one
obtains   \beq\label{intensite} {\bf I}^\varepsilon_\pm=-\varepsilon\hat{\bf
p}\,z_\pm \left[ 1+\frac{M^2}{2}  \int_{-1}^{1}{{\rm d}\cos\theta  {\cos\theta
(1-\cos\theta)\over (\omega_\pm-p{\rm cos}\theta)^2}}\right],\eeq  where we
have
neglected $p$ compared to $k$ in  evaluating   the second terms. It can be
verified that these expressions coincide with the velocity of the excitations
as
deduced from the dispersion relation $\omega_\pm(p)$. For example, for the hole
excitation,  we have  \beq \frac{d\omega_\pm(p)}{d{\bf
p}}=z_\pm\left(\frac{d\epsilon_p}{d{\bf p}} +\frac{\del\Sigma_u(\omega_\pm,
{\bf
p})}{\del{\bf p}}\right) \eeq with $z_\pm=1/\left(1-\del\Sigma_u(\omega,{\bf
p})/\del\omega\right)$, $\Sigma_u=a+bp$  denotes the mass operator in the
eigenspace corresponding to  $\varepsilon=1$, and ${\bf
I}_\pm^\varepsilon=-\varepsilon(d\omega_\pm/d{\bf p})$.
By carrying out  the integrals in
(\ref{intensite}), and using the results of section 2, one can rewrite the
currents as follows
\beq\label{intensity}
{\bf I}^\varepsilon_\pm=-\varepsilon\hat{\bf
p}\left[2z_\pm +\frac{\omega_\pm}{p}(1-2z_\pm)\right] \eeq
As we did earlier in the case of the charge, we can interpret
Eq.(\ref{intensity}) as the
sum of a ``direct'' contribution, that carried by the bare single particle,
i.e. $-\varepsilon\hat{\bf p} z_\pm$, and an ``induced'' contribution coming
from the hole-photon component of the states (\ref{statem0}). The hole photon
part of
the quantum states gives a contribution  to the current which is opposite to
that
of  the current carried by the bare particle. When $p\rightarrow 0$,  ${\bf
I}^\varepsilon_\pm=-\varepsilon\hat{\bf p}/3$.
Thus, at very low momentum, the sign of the current is the same as for the
corresponding bare particle: the negative  group velocity on the abnormal
branch
with $\omega>0$ simply reflects the fact that the corresponding bare particle
is
a {\it positron\/} state dressed by the medium, in spite of the fact that it
has
positive $\omega$. When $p$ becomes large, the current goes to $1$ on both
positive $\omega$ branches. However this is achieved in different ways on the
normal and the abnormal branch.On the normal branch, the residue $z_+\to 1$
and,
at large $p$ the current is carried entirely by the bare hole. On the abnormal
branch, $z_-\to 0$, and the current is no longer carried by the positron
component, but only by the hole-photon component. This explains  the minimum
(for $\omega>0$) in the abnormal branch of the dispersion relation. Indeed, as
we
have seen, at small $p$ the current is mostly that of the bare
positron. As $p$ increases, the weight of the hole-photon component
increases until it completely dominates. At this point, the current is that of
the hole in the hole-photon state, and we have seen that the states with
$\omega\sim p$ correspond to holes with momentum parallel  to ${\bf p}$, and
therefore to current opposite to that of the positron.

We now turn to the case of a massive fermion at rest ($p=0$). Then \beq \bar
u_{\lambda 0} u_{\nu,{\bf k}}=\frac{1}{\sqrt 2}\phi^\dagger_{\lambda }\phi_{\nu
\hat k} \qquad\qquad \bar v_{\lambda 0} u_{\nu,{\bf k}}=\frac{-\nu}{\sqrt
2}\phi^\dagger_{\lambda}\phi_{\nu\hat k}  \eeq The eigenvectors are therefore
\beq
\label{statep0} \ket{\Psi_0^u}_{\pm,\lambda}&=&\sqrt{z_\pm}\left(
b_{\lambda,0}+
\frac{g}{\omega_\pm\sqrt{2}}\sum_{\nu,{\bf k}}
\frac{\phi^\dagger_{\lambda}\phi_{\nu\hat k}}
{\sqrt{2\omega_k\Omega}}b_{\nu,{\bf
k}}a^\dagger_{\bf k}\right)\ket{\Phi_0}  \nonumber\\
\ket{\Psi_0^v}_{\pm,\lambda}&=&\sqrt{z_\pm}\left( d^\dagger_{\lambda,0}+
\frac{g}{\omega_\pm\sqrt{2}}\sum_{\nu,{\bf k}}
\frac{-\nu\,\phi^\dagger_{\lambda}\phi_{\nu\hat k}}
{\sqrt{2\omega_k\Omega}}b_{\nu,{\bf k}}a^\dagger_{\bf k}\right)\ket{\Phi_0}
\eeq
The normalisation constant $z_\pm$ can be calculated directly, as for $m=0$.
However, both helicity states now contribute to the sum and one must use
$\sum_{\nu}\left|\phi^\dagger_{\lambda}\phi_{\nu\hat k}\right|^2=1$. This gives
immediately  $z_\pm^{-1}=1+M^2/\omega_\pm^2$, in agreement with
Eq.(\ref{residp0}). The orthogonality of states with different $\omega$ can
also be
checked explicitly. One
could also repeat here the discussion given above in the case $m=0$
concerning the decoupling of the
single particle excitation which occurs when  $m\simge M$.

A remarkable fact in Eq.(\ref{statep0}) is that $\omega$ factorizes out of the
hole-photon component, so that, up to a trivial multiplication factor
$1/\omega_\pm$,
the same  coherent superposition of hole-photon states contributes to
both the  normal and the abnormal branch. Thus for example, we may write the
eigenstate  corresponding to the normal branch $\omega_+$
 in the form \beq \ket{\Psi_0^u}_{+,\lambda}&=&\sqrt{z_+}
b_{\lambda,0}\ket{\Phi_0} +\sqrt{1-z_+}\ket{\Phi_1} \eeq where $\ket{\Phi_1}$
is
the collective state. The  eigenstate corresponding to the abnormal branch
$\omega_-$ is also   a linear combination of $b_{\lambda,0}\ket{\Phi_0}$ and
$\ket{\Phi_1}$.  Since $\omega_+\not= \omega_-$, the eigenstates are orthogonal
to each other  and we may write, up to a phase,
\beq
\ket{\Psi_0^u}_{-,\lambda}&=&\sqrt{1-z_+}b_{\lambda,0}\ket{\Phi_0}
-\sqrt{z_+}\ket{\Phi_1}
\eeq
and thus $z_{-}=1-z_+$. The explicit construction of the eigenstates
thus explains naturally why the residues add up to unity when $p=0$ (see
section 2). On the other hand, for a massless fermion with ${\bf
p}\not= 0$, Eq.(\ref{statem0}) shows that the hole-photon component is
different
for the two branches and the previous  argument cannot be used. Indeed, the
sum of the residues is smaller than unity  in this case, some of the specral
weight being displaced in the region $-p<\omega<p$.

Let us finally consider the general case when $m$ and $p$ are both different
from zero. Then the solutions $\chi$ of Eq.(\ref{intDirac}) are linear
superpositions
of the solutions $w_{\lambda,{\bf p}}$ of the free Dirac hamiltonian. For
instance, a solution with $\varepsilon=+1$ and  helicity $\lambda$
can be written as a superposition of $u_{\lambda,{\bf p}}$ and $v_{\lambda,{\bf
p}}$
defined in Eq.(\ref{chiralpw}):
\beq
\chi = \sqrt{1-r_\pm}\ u_{\lambda,{\bf p}} +
{\rm e}^{i\varphi}\sqrt{r_\pm}\ v_{\lambda,{\bf p}}
\eeq
where $r_+ $ (resp. $r_-$) measures the mixing between the hole state
$b_{\lambda,{\bf p}}\ket{\Phi_0}$  and the positron state
$d^\dagger_{\lambda,-{\bf p}}\ket{\Phi_0}$ on the normal  (resp. abnormal)
branch.  For a state with
$\varepsilon=-1$, $u$ and $v$ must be interchanged in the  latter equation.
Thus the eigenstates are of the form
\beq
\label{mixedstate}
\ket{\Psi_{-\bf p}^u}_{\pm,\lambda}=\sqrt{z_\pm}\left(
\sqrt{1-r_\pm}\ b_{\lambda,{\bf p}}\ket{\Phi_0}+
\sqrt{r_\pm}\ d^\dagger_{\lambda,{-\bf p}}\ket{\Phi_0}\right)
+ \sqrt{1-z_\pm}\ket{\Phi_1}
\eeq
if $\varepsilon=+1$, and $b$ and $d^\dagger$ must be interchanged if
$\varepsilon=-1$. In Eq.(\ref{mixedstate}),
$z_\pm$ is the residue at the quasiparticle pole and $\ket{\Phi_1}$ is a
linear  superposition of  states
$b_{\nu,{\bf p}+{\bf k}}a^\dagger_{\bf k}\ket{\Phi_0}$
as in Eqs.(\ref{statem0}) and (\ref{statep0}).
 From Eq.(\ref{mixedstate}) we deduce that the probability
that a hole state $b_{\lambda,{\bf p}}\ket{\Phi_0}$ be
found in the eigenstate $\ket{\Psi_{-\bf p}^u}_{\pm,\lambda}$
(resp. $\ket{\Psi_{-\bf p}^v}_{\pm,\lambda}$) is $z_\pm (1-r_\pm)$
(resp. $z_\pm r_\pm$).

The mixing coefficients $r_\pm$ can be  calculated easily by noting that the
matrices $\gamma^0$ and  $\bfalpha\cdot\hat{\bf p}$ appearing in
Eqs.~(\ref{freeDirac}) and
(\ref{intDirac})   obey the same commutation relations as two arbitrary Pauli
matrices,
say $\sigma_1$ and $\sigma_2$. On may then associate a
solution to Eq.~(\ref{freeDirac}) to a spinor pointing in a direction
making an angle $\gamma$ with the $1$-axis, with ${\rm tan}\gamma=p/m$,
and the corresponding solution to Eq.~(\ref{intDirac}) to a spinor pointing in
a direction  making an angle $\gamma'$ with the $1$-axis, with ${\rm
tan}\gamma'=(1+b)p/m$.  The overlap between the two spinors is then  $\left|
w^\dagger w_{\lambda,{\bf p}}\right| =\sqrt{1-r_\pm}={\rm
cos}((\gamma-\gamma')/2)$ or,  explicitly,
\beq \label{mixing}
r_\pm={1\over 2}\left[ 1-{1+(1+b)p^2/m^2\over\sqrt{1+p^2/m^2}
\sqrt{1+(1+b)^2p^2/m^2}}\right] .
\eeq
It depends on $\omega$ through $b$. For the normal branch (resp. the
abnormal branch), $b$ varies from
$0$ to $-1/3$ (resp. from $-\infty$ to $-1/3$) when the chemical potential
increases from $0$ to $+\infty$. Thus the bounds on the
mixing coefficient $r_\pm$ are
\beq
0 < r_+ < {1\over 2}\left[ 1-{1+(2/3)p^2/m^2\over\sqrt{1+p^2/m^2}
\sqrt{1+(4/9)p^2/m^2}}\right] < r_- <
{1\over 2}\left[ 1+{p/m\over\sqrt{1+p^2/m^2}}\right].
\eeq
The maximum value of $r_+$ for the normal branch is reached when
$p/m=\sqrt{3/2}$ and infinite density:
$r_+^{\rm max}=1/2-\sqrt{6}/5\simeq 10^{-2}$,
which is rather low. On the other hand, for the abnormal branch the
mixing can exceed $1/2$ if the density is not too high; however, the
residue $z_-$ of the abnormal branch is then very small, so that the product
$z_- r_-$, which is the probability that a hole state $b_{\lambda,{\bf
p}}\ket{\Phi_0}$ be  found in the state $\ket{\Psi^v}_{-,\lambda}$,
remains small,  at most of the order of $10^{-2}$.

This mixing between holes and positrons can be easily understood
in the ultra--high density limit $M\gg m,p$. In this limit
Eq.(\ref{branches}) gives $\omega\rightarrow\pm M$ for both
values of $\varepsilon$ (see Fig.1a). Thus there is a degeneracy at
$\omega=M$ (resp. $-M$) between the branch $\omega_+$ (normal hole branch)
and the branch $-\omega_-$ (abnormal positron branch).
This degeneracy is lifted by terms proportional to $p$ and $m$, which
act like a perturbation in the eigenvalue equation (\ref{intDirac}) and
therefore
induce mixing between the two branches.
Writing Eq.(\ref{intDirac}) in the form $(H_0(\omega)+V)\chi=\omega\chi$ with
$H_0(\omega)=a\simeq M^2/\omega$ and $V=(2/3)\bfalpha\cdot{\bf p}+\gamma^0 m$
(we have used the fact that $b\rightarrow -1/3$ for $|\omega|\gg p$),
and expanding around $\omega=M$ one gets
\beq
\left[ \left({dH_0\over d\omega}\right)_{\omega=M}(\omega-M)+V\right]\chi=
(\omega-M)\chi.
\eeq
Since $(dH_0/d\omega)_{\omega=M}=-1$, the problem reduces to diagonalizing
$V$. In particular, one gets immediately the eigenvalues
$\omega=M\pm\sqrt{p^2/9+m^2/4}$. Note that $r_+=r_-$ in this limit,
and that the mixing vanishes if either
$m=0$ or $p=0$, as expected.

The calculation of the eigenstates that we have performed in this section,
which
is equivalent to solving Dyson's equation for the poles of the Green function
in the ultrarelativistic limit, assumes that the modifications of the ground
state can be
ignored. More generally, the solution of Dyson's equation for the full one loop
self energy is equivalent to the following eigenvalue problem
\beq
\label{eigenst1}
&&(\omega-\epsilon_{\lambda, p})X_{\lambda,{\bf p}}
-\sum_{k \nu}\frac{g}{\sqrt{2\omega_k\Omega}}
\bar w_{\nu, {\bf p}+{\bf k}}w_{\lambda,{\bf p}} Y_{\nu,{\bf p},{\bf k}}
-\sum_{k \nu}\frac{g}{\sqrt{2\omega_k\Omega}}
\bar w_{\nu, {\bf p}+{\bf k}}w_{\lambda,{\bf p}} Z_{\nu,{\bf p},{\bf
k}}=0\nonumber\\
&&-\frac{g}{\sqrt{2\omega_k\Omega}}\sum_\lambda
\bar w_{\nu, {\bf p}+{\bf k}}w_{\lambda,{\bf p}} X_{\lambda,{\bf p}}
+(\omega-\epsilon_{\nu,{\bf p}+{\bf k}}+\omega_k)Y_{\nu,{\bf p},{\bf
k}}=0\nonumber\\
&&-\frac{g}{\sqrt{2\omega_k\Omega}}\sum_\lambda
\bar w_{\nu, {\bf p}+{\bf k}}w_{\lambda,{\bf p}} X_{\lambda,{\bf p}}
+(\omega-\epsilon_{\nu,{\bf p}+{\bf k}}-\omega_k)Z_{\nu,{\bf p},{\bf k}}=0
\eeq
where, in the first equation the sum runs only over the states such that
$\epsilon_{\bf
p+k}<\mu$ for the $Y$ amplitudes, and such that $\epsilon_{\bf
p+k}>\mu$ for the $Z$ amplitudes. That this system of equations is equivalent
to Dyson's
equation  can be easily verified by eliminating the amplitudes $Y_{\nu,{\bf
p},{\bf k}}$ and
 $Z_{\nu,{\bf p},{\bf k}}$ with the help of the last two equations, and
comparing the
resulting equation for $X_{\lambda,{\bf p}}$ with Dyson's equation. This
comparison allows us
to identify the mass operator, and one recovers indeed Eq.~(\ref{fullsigma}).
The equations (\ref{eigenst1}) take
into account the modifications of the Fermi sea due to interactions, at this
order of
perturbation theory. In particular, because of the interactions, single
particle levels are not
fully occupied below the Fermi level, and there are non vanishing amplitudes of
the form \beq
Z_{\nu,{\bf p},{\bf k}}=\bra{\Psi_{\nu,{\bf p},{\bf k}}}b^\dagger_{\nu,{\bf
p}+{\bf k}}a_{\bf
k}\ket{\Psi_0}, \eeq
where $\ket{\Psi_0}$ and $\bra{\Psi_{\nu,{\bf p},{\bf k}}}$ denote true
eigenstates,
that is including perturbative  corrections. As for the amplitudes $X$ and $Y$
they are given
as before by
\beq
X_{\nu,{\bf k}}=\bra{\Psi_{\nu,{\bf p}}}b_{\nu,{\bf p}}\ket{\Psi_0}\qquad\qquad
Y_{\nu,{\bf p},{\bf k}}=\bra{\Psi_{\nu,{\bf p},{\bf k}}}b_{\nu,{\bf p}+{\bf k}}
a^\dagger_{\bf k}\ket{\Psi_0},
\eeq
but now, the state $\ket{\Psi_0}$ contains perturbative corrections, i.e.
it differs from the unperturbed Fermi sea $\ket{\Phi_0}$. It is easily seen
on Eq. (\ref{eigenst1}) that in the ultrarelativistic limit, and for soft
modes, the amplitudes  $Z_{\nu,{\bf p},{\bf k}}$  become negligible as
compared to the $Y_{\nu,{\bf p},{\bf k}}$ ones.

\section{Discussion}

We have, in the main text, described a model plasma composed of Dirac fermions
interacting
with a scalar field. Most of our results extend however to
a dense system of electrons (see Appendix C),
or to dense quark matter. In such
 high density systems, particle masses are negligible in a first
 approximation, and the only energy scale is provided by
the
  chemical potential, $\mu$. Most particles in such a system
   have momenta of order $\mu$.
Long wavelength
excitations develop on the  scale $g\mu$, and
have a  collective character.  Bosonic modes, not discussed in
this paper, correspond to the  usual plasmons involving
particle-hole excitations localized in the vicinity of the
Fermi surface. We have concentrated here on  fermionic modes,
which can be excited by processes adding small momentum positrons or
removing electrons at the bottom of the Fermi sea.

The physical picture which emerges from our calculation is quite
in line with the work of Ref.\cite{QED} where the
equations of motions describing the long wavelength
excitations of an ultrarelativistic plasma at high
temperature are reduced to a set of coupled mean field and
 kinetic equations incorporating the
dominant medium effects as ``polarization'' phenomena. In
either case of large temperature or large chemical potential,
the polarization is entirely due to the
hard particles with momenta of order $T$ or $\mu$.
In the zero temperature case, the dominant process
is that in which a hard electron
annihilates with the soft hole, or the soft positron, and
turns into a hard photon. Thus, the quantum numbers of the
hard particles fluctuate, with a typical period of order
$1/g\mu$. This is accompanied by a
corresponding oscillation in the number of soft
particles, which is interpreted as an oscillation of the
average soft fermionic field in Ref.\cite{QED}.
The present study allows us to give a simple illustration
of this in terms of the quantum states that we have
constructed. Consider indeed a perturbation which consists
in adding soft positrons to, or creating soft holes in, the
unperturbed Fermi sea at time $t=0$, say. The system will
then evolve into a state $\ket{ \Phi(t)}$ which will contain components
on the states $\ket{\Psi^\varepsilon}_\pm$ constructed in section 4:
\beq
\ket{\Phi(t)}=\ket{\Phi_0}+\alpha(t) \ket{\Psi^\varepsilon}_\pm\eeq
 where $\alpha(t)$ is a small time dependent amplitude.
The expectation value of the fermionic field in such a
state is proportional to the amplitudes
$X_{\lambda,{\bf p}}$ introduced in section 4, while the
amplitudes $Y_{\lambda,{\bf p},{\bf k}}$ contribute to the ``induced
source'' for the fermion field.

The explicit construction of quasiparticle states in section
4 clearly shows how new modes emerge: the free electron hole is coupled
to a large number of states which do not interact together and
have an energy comparable to that of the unperturbed hole excitation.
To outline this structure, let us consider a simplified situation where a
quantum state $\ket{0}$ with energy $E_0$ (corresponding to
$\epsilon_p$ in the previous section) is coupled with equal strength
$V$ to a large number $N$ of states $\ket{i}$, uniformly distributed
throughout the energy interval
$[-\Delta E/2,\Delta E/2]$.
In the calculation performed in section 4, $\Delta E=2p$,
$N$ is of order $\Omega\mu^3$ in a
volume $\Omega$ and $V$ is of order $g/\sqrt{\Omega\mu}$, i.e. $V\sqrt{N}\sim
g\mu$.
The equation determining the modes $\omega$ is
\beq
\omega-E_0=-{NV^2\over \Delta E}
\log\left|{\omega+\Delta E/2\over\omega-\Delta E/2}\right|
\eeq
A substantial energy shift and/or a new quasiparticle will occur only if
$V\sqrt{N}$
is at least of the same order of magnitude as both
$E_0$ (which is the difference between the energy of $\ket{0}$
and that of the centroid of the states $\ket{i}$ to which $\ket{0}$
couples)
and $\Delta E$ (the energy interval over which the
states $\ket{i}$ are spread). For the calculation  of section 4,
this is equivalent to saying that $g\mu$ must be at least as large as $m$ and
$p$.

One can then wonder whether a similar situation could occur
in other systems. In fact, a
structure similar to that studied here
has been found in the study of the electron gas at
intermediate densitites\cite{Hedin67,Mahaux85}. There, the coupling of an
electron hole
to collective plasmons gives rise to a peak in the imaginary part
of the hole self-energy which is sufficiently strong  to produce
a new quasiparticle, dubbed the ``plasmaron''. However, in the case of the
non relativistic electron gas, the validity of the perturbative
approach is less clear, and it was indeed found that vertex
corrections could very well suppress the structure\cite{Minnhagen74}.

\vspace{1cm}
\noindent{\bf Acknowledgements}

We are grateful to E.~Iancu for a critical reading of the
manuscript and  we thank him for  many useful remarks. One of us (JPB) wishes
to thank B.~Friman for mentioning the works on the plasmaron
quoted in refs. \cite{Hedin67,Minnhagen74,Mahaux85}

\renewcommand{\theequation}{A.\arabic{equation}}
\newpage
\setcounter{equation}{0}

\appendix
\noindent {\bf APPENDIX A~: The vacuum part of the self--energy}

At zero temperature and zero chemical potential, the mass
operator  $\Sigma_0({\bf p},\omega)$  may be written as
\beq
\Sigma_0({\bf
p},\omega)=a_0(p^2,p_0)\gamma_0+b_0(p^2)\bfgamma\cdot
{\bf p}+c_0(p^2)
\eeq
where, because of Lorentz invariance, the functions
$a_0/p_0,b_0$ and $c_0$ depend only on the 4-momentum
squared, $p^2=p_0^2-{\bf p}^2$, and
$b_0(p^2)=-a_0(p^2,p_0)/p_0$. We shall limit the
discussion to the case ${\bf p}=0$ (the result for
${\bf p}\not= 0$ may be obtained from Lorentz invariance).
The expressions
for $a_0$ and $c_0$ may be read off Eqs.~(\ref{sigma0}), and
they involve ultraviolet diverging integrals over
momenta. Adding the usual counterterms of mass and wave
function renormalization, we write the renormalized
inverse propagator as
\beq\label{renorm}
{G_R^{-1}(\omega )\over Z_2}&=&G_0^{-1}(\omega )
+\Sigma_0(\omega)-\delta m\nonumber\\
&=&\left( -\omega+a_0(\omega)\right)\gamma^0+\left(
m+c_0(\omega)-\delta m\right)
 \eeq
The counterterms are fixed by the on--shell renormalization
conditions \beq
\delta m&=& a_0(m)+c_0(m)\nonumber\\
Z_2^{-1}-1&=&1-{\partial
a_0\over\partial\omega}\biggr\vert_{\omega=m} -{\partial
c_0\over\partial\omega}\biggr\vert_{\omega=m}.  \eeq
In doing the calculations of $Z_2$, it is necessary to
give the photon a small mass $\lambda$ to avoid the infrared
divergence.  The Green function can
finally be written in the form
\beq G_R^{-1}(\omega )=\left(
-\omega +a_R(\omega )\right)+m+c_R(\omega),
\eeq
where $a_R(\omega)$ and $c_R(\omega)$
are the renormalized  contributions of the vacuum to the
mass operator, which must be added to $a_1(\omega)$ and
$c_1(\omega)$ in the finite density
case:
\beq\label{sigmavac} a_R(\omega )&=&{g^2\over
16\pi^2}\left[ {\omega^2-m^2\over 2\omega} \left(
1+{\omega^2+m^2\over\omega^2}\ln\left\vert 1-{\omega^2\over
m^2} \right\vert\right) -\omega\left( 5+4 \ln {\lambda\over
m}\right)\right]\nonumber\\ c_R(\omega )&=&{g^2m\over
16\pi^2}\left[ \left(1-{m^2\over\omega^2} \right)
\ln\left\vert 1-{\omega^2\over m^2} \right\vert +\left( 5+4
\ln {\lambda\over m}\right)\right] ,  \eeq
and we have kept only the dominant terms in $\lambda/m$ as
$\lambda\to 0$.

Note that the renormalization does not affect the
imaginary part which is finite. Note also that in
calculating the frequencies of the modes at high density, one can ignore
$a_R$ and $c_R$. Indeed, when $\omega\sim g\mu$, their contribution is of order
$g^3\mu$ while that of  $\Sigma_1$ is of order
$g\mu$ . Thus, in particular,  the term in
$\lambda$ which is infrared
divergent and, in gauge theories, gauge dependent, does
not contribute in the high density limit.
Presumably it would disappear in a consistent calculation of
quasiparticle energies at order $g^3\mu$, i.e. beyond one-loop .

\newpage
\renewcommand{\theequation}{B.\arabic{equation}}
\setcounter{equation}{0}

\appendix
\noindent {\bf APPENDIX B~: Exact expressions for the one--loop mass operator
at $T=0$}
\def\kF{k_F}

In this appendix we give the exact expression of the mass operator
(\ref{fullsigma}), valid
 when either $p$ or $m$ vanishes.

If $p=0$, the finite density contribution to the mass operator can be written
\beq \label{smfini}
\Sigma_1({\bf p}=0,\omega )=a_1(\omega )\gamma^0+c_1(\omega )
\eeq
The real parts of $a_1$ and $c_1$  are
\beq
\label{mfini}
{\rm Re}\left[ a_1(\omega )\right] &=&{g^2m\over 16\pi^2}
\Biggl[ {\sqrt{\mu ^2-m^2}\over m}\left(
1+{m^2\over \omega^2}+{\mu\over\omega}\right) +{m^3\over\omega^3}\cosh ^{-1}
\left( {\mu\over m}\right) \nonumber\\
& &\phantom{{g^2m\over 16\pi^2}\Biggl[\qquad\qquad}
+{\omega\over 2m}\left( 1-{m^4\over\omega^4}\right)
\ln \left\vert {\omega -\mu -k_F \over \omega -\mu +\sqrt
{\mu^2-m^2}} \right\vert \Biggr] \nonumber\\
{\rm Re}\left[ c_1(\omega )\right] &=&{g^2m\over 8\pi^2}\Biggl[
{k_F\over\omega}+
{m^2\over\omega^2}\cosh^{-1}\left( {\mu\over m}\right) \nonumber\\
&&\phantom{{g^2m\over 16\pi^2}\Biggl[\qquad\qquad}
+{1\over 2}\left( 1-{m^2\over\omega^2}\right)
\ln \left\vert {\omega -\mu -k_F \over \omega -\mu +\sqrt
{\mu^2-m^2}} \right\vert \Biggr]
\eeq
with $\kF=k_F$ the Fermi momentum.
The imaginary parts are (including the vacuum contribution)
\beq
\label{imfini}
{\rm Im} \left[ a_1(\omega)\right] &=&-\pi{g^2\over
16\pi^2}\,{m^4-\omega^4\over 2\omega^3}\nonumber\\
{\rm Im} \left[ c_1(\omega)\right] &=&-\pi{g^2m\over
16\pi^2}\,{m^2-\omega^2\over \omega^2}
\eeq
if
\beq
\label{condit}
\omega <-m\ \ \ {\rm or}\ \ \ \mu-k_F<\omega <m
\eeq
and $0$ otherwise. Note that this equation agrees with
Eq.(\ref{imultra}) when we set $p=0$ in the latter.
We note the relation
${\rm Im}(a+c)/{\rm Im}(a-c)=(\omega+m)^2/(\omega-m)^2$,
identical to that found at high $T$
and $\mu=0$ \cite{BBS92}. The imaginary part is independent of $\mu$,
except for the domain where it is non vanishing.
In the high density limit, this domain is
 the interval $m^2/(2\mu)<\omega <m$; the limit $m\to 0$ is
thus singular.

If $m=0$, $k_F=\mu$, the finite density contribution to the mass operator can
be written
\beq
\Sigma_1({\bf p},\omega )=a_1({\bf p},\omega )\gamma^0+b_1({\bf p},\omega )
\mbox{\boldmath$\gamma$}\cdot{\bf p}
\eeq
The functions $a_1$ and $b_1$ can be calculated analytically by
choosing  $({\bf p}-{\bf k})$ as integration variable
in Eq.~(\ref{sigma1}). The real parts are:
\beq\label{pfini}
{\rm Re}\left[ a_1(p,\omega )\right] &=
&{g^2\over 32\pi^2p}\Biggl\{ p\mu
+{(2\mu-\omega-p)(2\mu+\omega+p)\over 4}\ln
\left\vert {\omega +p\over 2\mu-\omega -p}\right\vert\nonumber\\
&&\phantom{{g^2\over 32\pi^2p}\Biggl\{ p\mu }
-{(2\mu-\omega+p)(2\mu+\omega-p)\over 4}\ln
\left\vert {\omega -p\over 2\mu-\omega +p}\right\vert \Biggr\} \nonumber\\
& & \phantom{aaaaaaaaaaaa}\nonumber\\
{\rm Re}\left[ b_1(p,\omega )\right] &=& {g^2\over 16\pi^2p^2}\Biggl\{ \mu^2
+{2\mu-\omega-p\over 4p}\left[ \omega^2-p^2-{\omega\over 2}
(2\mu+\omega +p)\right]\ln\left\vert {\omega +p\over
2\mu-\omega -p}\right\vert \nonumber\\
&& \phantom{{g^2\over 16\pi^2p^2}\Biggl\{ \mu^2}
-{2\mu-\omega+p\over 4p}\left[ \omega^2-p^2-{\omega\over 2}
(2\mu+\omega -p)\right]\ln\left\vert {\omega -p\over
2\mu-\omega +p}\right\vert\Biggr\} \nonumber\\
&&
\eeq
As to the imaginary parts, several cases must be distinguished:
\beq\label{impfini}
III:\qquad{\rm Im}\left[a(p,\omega)\right] &=& \phantom{-}\pi{g^2\over 32\pi^2}
\omega\nonumber\\
{\rm Im}\left[b(p,\omega)\right] &=& -\pi{g^2\over 32\pi^2} \nonumber\\
II:\qquad{\rm Im}\left[a(p,\omega)\right] &=& -\pi{g^2\over 32\pi^2}
{(2\mu-\omega-p)(2\mu+\omega+p)\over 4p}\nonumber\\
{\rm Im}\left[b(p,\omega)\right] &=& -\pi{g^2\over 32\pi^2}
{2\mu-\omega-p\over 2p^3}\left[ \omega^2-p^2-{\omega\over 2}
(2\mu+\omega +p)\right]\nonumber\\
IV:\qquad{\rm Im}\left[a(p,\omega)\right] &=& {\rm Im}\left[b(p,\omega)\right]
= 0 \nonumber\\
Ia:\qquad{\rm Im}\left[a(p,\omega)\right] &=& \phantom{-}\pi{g^2\over 32\pi^2}
{(2\mu-\omega-p)(2\mu+\omega+p)\over 4p}\nonumber\\
{\rm Im}\left[b(p,\omega)\right] &=& \phantom{-}\pi{g^2\over 32\pi^2}
{2\mu-\omega-p\over 2p^3}\left[ \omega^2-p^2-{\omega\over 2}
(2\mu+\omega +p)\right]\nonumber\\
Ib:\qquad{\rm Im}\left[a(p,\omega)\right] &=& -\pi{g^2\over 32\pi^2}
\omega \nonumber\\
{\rm Im}\left[b(p,\omega)\right] &=& \phantom{-}\pi{g^2\over 32\pi^2}
\eeq
where the various regions are indicated on Fig.10 ($III:\omega<-p;
II:-p<\omega<\min(p,2\mu-p);IV:\min(p,2\mu-p)<\omega<\max(p,2\mu-p);
Ia: \max(p,2\mu-p)<\omega<2\mu+p;Ib:\omega>2\mu+p$). These expressions include
the contribution of the vacuum to the imaginary
parts.
One checks easily that these equations give the same result in the limit
$p\to 0$ as Eqs.(\ref{mfini}) and (\ref{imfini}) when $m\to 0$.
Furthermore, if $\mu\gg\omega, p$, one recovers Eqs.(\ref{abcultra}) and
Eq.(\ref{Landau}). Note that the imaginary part is everywhere negative.

\newpage
\renewcommand{\theequation}{C.\arabic{equation}}
\setcounter{equation}{0}

\def\bfepsilon{\mbox{\boldmath$\epsilon$}}
\appendix
\noindent {\bf APPENDIX C~: QED}

The results obtained in sections 2 and 4 in the case where fermions
are coupled to a massless scalar field can be easily generalized to
an electromagnetic coupling. The corresponding modifications are
summarized in this appendix.

In the Coulomb gauge, the electromagnetic field is decomposed in
\beq\label{Vector}
A^\mu(x)=\frac{1}{\sqrt{\Omega}}\sum_{\bf k}\sum_{\rho=1,2}
\frac{1}{\sqrt{2\omega_k}}\left(
e^{-ik\cdot x}\, \epsilon^\mu_{\rho,{\bf k}}a_{\rho,{\bf k}}+
e^{ik\cdot x}\, \epsilon^{\mu,*}_{\rho,{\bf k}}a^\dagger_{\rho,{\bf k}}
\right),
\eeq
where $\rho=1,2$ is a polarization index and
$\epsilon^\mu_{\rho,{\bf k}}$ are normalized polarization vectors~:
$\epsilon^0_{\rho,\bf k}=0$, $\bfepsilon_{\rho,\bf k}\cdot{\bf k}=0$ and
$\bfepsilon_{\rho,\bf k}\cdot\bfepsilon_{\rho',\bf k}=\delta_{\rho,\rho'}$.
The interaction hamiltonian is now
\beq
H_{int}&=&g\int{\bar\psi(x)\slashchar{A}(x)\psi(x)}d^3x\nonumber\\
&=&\frac{g}{\sqrt{\Omega}}\sum_{{\bf p},{\bf
k},\lambda,\nu,\rho}\frac{1}{\sqrt{2\omega_k}}
\left( \bar w_{\nu,{\bf p}+{\bf k}}\slashchar{\epsilon}_{\rho,{\bf
k}}w_{\lambda,{\bf p}}
a_{\rho,\bf k}
+\bar w_{\nu,{\bf p}+{\bf k}}\slashchar{\epsilon}^*_{\rho,{\bf
k}}w_{\lambda,{\bf p}}
a^\dagger_{\rho,-\bf k}\right)
b^\dagger_{\nu,{\bf p}+{\bf k}}b_{\lambda,{\bf p}}
\nonumber\\
\eeq
and the result of the one loop calculation may be obtained,
in the ultrarelativistic limit, by diagonalizing $H$ in the basis
\beq\label{vstates}
b_{\lambda,{\bf p}}\ket{\Phi_0}\qquad\qquad b_{\nu,{\bf p}+{\bf
k}}a^\dagger_{\rho,\bf k}\ket{\Phi_0}.
\eeq
The corresponding amplitudes will be denoted by $X_{\lambda,{\bf p}}$ and
$Y_{\lambda,\rho,{\bf p},{\bf k}}$. The eigenvalue equation takes the form
(\ref{elim1}) with
\beq\label{vnewsigma}
M({\bf p},\bar\omega)=g^2\int\frac{d^3k}
{(2\pi)^3}\frac{1}{2\omega_k}\bigg\{
\frac{\slashchar{\epsilon}^*_{\rho,\bf k}\Lambda^+_{{\bf p}+{\bf k}}\gamma^0
\slashchar{\epsilon}_{\rho,\bf k}f_{{\bf p}+{\bf k}}}
{-\bar\omega-\epsilon_{{\bf p}+{\bf k}}+\omega_k}
+\frac{\slashchar{\epsilon}^*_{\rho,\bf k}\Lambda^-_{{\bf p}+{\bf k}}
\gamma^0\slashchar{\epsilon}_{\rho,\bf k}}
{-\bar\omega+\epsilon_{{\bf p}+{\bf k}}+\omega_k}
\bigg\} .
\eeq
In the ultrarelativistic limit ($k\gg m,p$),
the term proportional to $m$ in $\Lambda^\pm({\bf p}+{\bf k})$ can be
neglected, and ${\bf p}+{\bf k}$ replaced by ${\bf k}$.
Then, since $\bfepsilon_{\rho,\bf k}\cdot{\bf k}=0$,
$\slashchar{\epsilon}_{\rho,\bf k}=-\bfgamma\cdot\bfepsilon_{\rho,\bf k}$
anticommutes with $\bfgamma\cdot{\bf k}$ and thus with
$\Lambda^\pm_{\bf k}\gamma^0$. Then, using
$\sum_\rho{\slashchar{\epsilon}^*_{\rho,\bf k}\slashchar{\epsilon}_{\rho,\bf
k}}=-2$,
one gets
\beq
M({\bf p},\bar\omega)_{\rm Coulomb}=2M({\bf p},\bar\omega)_{\rm scalar}.
\eeq
Thus, the only difference with the Yukawa coupling is a global factor of two
in the mass operator, which corresponds to the fact that there are two
spin states for the bosons instead of one. The same result holds more
generally for the mass operator $\Sigma$ in the ultrarelativistic limit,
even at finite $T$, so that the results of section 2 are
also valid in the case of QED, if the definition of $M$ in Eq.(\ref{defM})
is replaced by
\beq
M={g\sqrt{2}\over 4\pi}\sqrt{\mu ^2+\pi ^2T^2}.
\eeq

Let us now construct the eigenstates explicitly. The difference with the scalar
field
lies in that the spinor overlap $\bar w_{\nu, {\bf p}+{\bf k}}w_{\lambda,{\bf
p}}$
in Eq.(\ref{eigenst}) must be replaced by
$\bar w_{\nu, {\bf p}+{\bf k}}\slashchar{\epsilon}^*_{\rho,\bf
k}w_{\lambda,{\bf p}}$.
If $m=0$,
\beq
\bar u_{\lambda,{\bf p}}\slashchar{\epsilon}^*_{\rho,\bf k} u_{\nu,{\bf k}}
=-\lambda\phi^\dagger_{\lambda\hat p}\bfepsilon_{\rho,\bf
k}\cdot\bfsigma\phi_{\nu\hat k}
\delta_{\nu,\lambda}
\ \ \
\bar v_{\lambda,{\bf p}}\slashchar{\epsilon}^*_{\rho,\bf k} u_{\nu,{\bf k}}
=-\phi^\dagger_{\lambda\hat p}\bfepsilon_{\rho,\bf k}\cdot\bfsigma\phi_{\nu\hat
k}
\delta_{\nu,-\lambda} .
\eeq
The electromagnetic interaction, unlike the
scalar coupling, conserves chirality. The eigenstates
are therefore
\beq\label{vstatem0}
\ket{\Psi^u_{-\bf p}}_{\pm,\lambda}&=&\sqrt{z_\pm}\left( b_{\lambda,{\bf p}}+
g\sum_{\rho,\bf k}\frac{1}{\sqrt{2\omega_k\Omega}}
\frac{-\lambda\,\phi^\dagger_{\lambda\hat p}
\bfepsilon^*_{\rho,\bf k}\cdot\bfsigma\phi_{\lambda\hat k}}
{\omega_\pm-p\ {\rm cos}\theta}
b_{\lambda,{\bf p}+{\bf k}}a^\dagger_{\rho,\bf k}\right)\ket{\Phi_0}
\nonumber\\
\ket{\Psi^v_{-\bf p}}_{\pm,\lambda}&=&\sqrt{z_\pm}\left(
d^\dagger_{\lambda,-{\bf p}}+
g\sum_{\rho,\bf k}\frac{1}{\sqrt{2\omega_k\Omega}}
\frac{-\phi^\dagger_{\lambda\hat p}
\bfepsilon^*_{\rho,\bf k}\cdot\bfsigma\phi_{-\lambda\hat k}}
{\omega_\pm-p\ {\rm cos}\theta}
b_{-\lambda,{\bf p}+{\bf k}}a^\dagger_{\rho,\bf
k}\right)\ket{\Phi_0}.\nonumber\\
\eeq
In order to calculate the normalization, the electric current, etc, one must
evaluate
the sum over polarization states $\sum_\rho {\left| \phi^\dagger_{\lambda\hat
p}
\bfepsilon^*_{\rho,\bf k}\cdot\bfsigma\phi_{\pm\lambda\hat k}\right|
}=1\mp\cos\theta$.
These expressions are similar to those obtained in the case of a Yukawa
coupling, except
for a factor of 2 which is absorbed in the definition of $M$. So the results
are
finally the same.

In the case ${\bf p}=0$, a similar calculation gives
\beq\label{vstatep0}
\ket{\Psi^u_{-\bf p}}_{\pm,\lambda}&=&\sqrt{z_\pm}\left( b_{\lambda,0}+
\frac{g}{\omega_\pm\sqrt{2}}\sum_{\nu,\rho,{\bf k}}
\frac{-\nu\,\phi^\dagger_{\lambda}\bfepsilon^*_{\rho,\bf
k}\cdot\bfsigma\phi_{\nu\hat k}}
{\sqrt{2\omega_k\Omega}}b_{\nu,{\bf k}}a^\dagger_{\bf k}\right)\ket{\Phi_0}
\nonumber\\
\ket{\Psi^v_{-\bf p}}_{\pm,\lambda}&=&\sqrt{z_\pm}\left( d^\dagger_{\lambda,0}+
\frac{g}{\omega_\pm\sqrt{2}}\sum_{\nu,\rho,{\bf k}}
\frac{-\phi^\dagger_{\lambda}\bfepsilon^*_{\rho,\bf
k}\cdot\bfsigma\phi_{\nu\hat k}}
{\sqrt{2\omega_k\Omega}}b_{\nu,{\bf k}}a^\dagger_{\rho,\bf
k}\right)\ket{\Phi_0}.
\eeq

\newpage

\centerline{\bf FIGURE CAPTIONS}
\vskip\baselineskip
\parindent 0pt
\parskip\baselineskip

\noindent{\bf Fig.1:}\par\noindent
The spectrum of fermionic excitation in an ultra\-relati\-vistic plasma for two
cases, $M=10m$ (Fig.1a) and $M=m$ (Fig.1b). The full lines correspond to
$\varepsilon=1$, the dashed lines to $\varepsilon=-1$. The lower part
of the figures displays the residues at the quasiparticle poles: $z_+$
is the  residue of the normal branches and $z_-$ is the residue of
the  abnormal branches.

\noindent{\bf Fig.2:} \par\noindent
Graphs corresponding to the three contributions labelled
(i),(ii) and (iii) in the expression (\ref{fullsigma}) of
the mass operator.

\noindent{\bf Fig.3:} \par\noindent
Energy domains where the imaginary part of $\Sigma$,
given by Eq.(\ref{fullsigma}), is non
vanishing. The three regions
(I), (II) and (III)  correspond respectively to the electron-photon
continuum, the hole-photon continuum and the positron-photon
continuum, and are the regions in which, respectively, the
contributions (i),(ii) and (iii) to ${\rm Im}\Sigma$ are non
vanishing.

\noindent{\bf Fig.4:} \par\noindent
Diagrams contributing to the second order correction to the
energy of a hole (i) with momentum $p<k_F$ , and a particle (ii) with
momentum $p>k_F$, and reflecting modifications of the Fermi sea. The
diagram (0) represents the second order correction to the energy of
the Fermi sea, i.e. $\Delta E_0^N$ (see text). It is the sum over
the fermion momenta in this diagram which is restricted by the
presence of the extra particle (in (ii)) or the extra hole (in (i)).

\noindent{\bf Fig.5:} \par\noindent
Schematic representation of some important excitations of the
ultrarelati\-vis\-tic
plas\-ma at zero temperature. (a) A hole at the bottom of the Fermi sea. (b) A
positron with $p=0$, i.e. a hole at the top of the Dirac sea. (c) A hole
left by a ``hard'' electron with momentum $p\simle k_F$. This hole is
accompanied by a hard photon (not drawn).

\noindent{\bf Fig.6:} \par\noindent
a) {\it solid lines\/}:
real part and imaginary part of the mass operator
for an eigenstate of $\gamma^0$ with the eigenvalue $\varepsilon=1$. The
calculations
are done using Eq.~(\ref{mfini}), Eq.~(\ref{imfini}) and Eq.~(\ref{condit})
(the
vacuum contribution to the real part is thus omitted). The imaginary part
vanishes stricly on this graph for $-m<\omega<0$ and for $\omega >m$.
It is small but not zero for $\omega<-m$.
The labels on the curves (5,10,20) are the corresponding values of
$\mu/m$.  In this figure as well as in Figs.~4 and 5,
$g^2/4\pi=1/137$.  {\it Vertical dash-dotted lines\/}: indicate
the value $\omega=\mu-k_F$ where the mass operator is divergent.
The other divergence, at $\omega=\mu+k_F$ is outside the
range of the abscissa. {\it Dashed line\/}: line
$\omega-m,$ whose intersection with the solid line
gives the elementary modes according to Eq.~(\ref{mmodes}).
\par\noindent
b) Same as a) for an eigenstate of $\gamma^0$ with the value
$\varepsilon=-1$. The dashed
line is now $\omega+m$ according to Eq.~(\ref{mmodes}).

\noindent{\bf Fig.7:} \par\noindent
a) {\it solid lines\/}:
real part and imaginary part of the mass operator
for an eigenstate of $\bfalpha\cdot\hat p$
with the eigenvalue $\varepsilon=1$. The calculations
are done using Eq.~(\ref{pfini}) and Eq.~(\ref{impfini}).
The imaginary part vanishes strictly on this graph for $\omega>p$.
It is small but non vanishing for $\omega< -p$.
The labels (10,20,40) on the curves are the corresponding values of
$\mu/p$.  {\it Dashed line\/}: line
$\omega-p,$ whose intersection with the solid line
gives the elementary modes according to Eq.~(\ref{pmodes}). \par\noindent
b) Same as a) for an eigenstate of $\bfalpha\cdot\hat{\bf p}$ with
the value $\varepsilon=-1$.

\noindent{\bf Fig.8:} \par\noindent
Same as Fig.7 with $p$ very close to the Fermi surface
($k_F/p=0.8,1,1.2$).

\noindent{\bf Fig.9:}\par\noindent
The dispersion relation for the hole excitation. Left-hand side:
no interaction. Right-hand side: with interaction. The shaded area
represents the most important part of the continuum of hole--photon
states to which  the hole couples.

\noindent{\bf Fig.10:}\par\noindent
The energy domains where the imaginary part of the mass operator is
non vanishing, in the limit $m=0$ (see Eq.(B.7)).

\end{document}